# Quantum mechanical 4-dimensional non-polarizing beamsplitter

Artem Kryvobok[1*], Alan D. Kathman[2]

***Abstract*— Some quantum optics researchers might not realize that classical electromagnetism predicts a $\pi$ phase shift between S- and P-polarized reflection and might think the reflection coefficients of the transverse modes are independent, or that such a phase shift has no measurable consequences. In this paper we discuss theoretical grounds to define elements of a 4x4 matrix to represent the beamsplitter, accurately accounting for transverse polarization modes and phase relations between them. We also provide experimental evidence confirming this matrix representation. From scientific point of view the paper addresses a non-trivial equivalence between the classical fields Fresnel formalism and the canonical commutation relations of the quantized photonic fields. That the formalism can be readily verified with a simple experiment provides further benefit. The beamsplitter expression derived can be applied in the field of quantum computing.**

## I. Introduction

QUANTUM optical derivation of a 4-dimensional scattering matrix coefficients for a non-polarizing lossless beamsplitter is discussed in this paper. With the derivation we demonstrate that imposing the reciprocity relations alone is insufficient for the correct matrix formulation. We propose imposing the canonical commutation relations applied to the multimode (spatial and transverse) output fields. The commutators method appear providing correct and sufficient condition to deduce the matrix. The acquired coefficients appear to correspond exactly to what Fresnel amplitudes would be equivalent to in our setup. We regard this as an interesting outcome since the two theories involve completely different physics.

The derived 4x4 scattering matrix spans 4-dimensional vector space with two spatial and two orthogonal polarization modes and represents an exact phase distribution for resulting superposed fields. Such a matrix and the derivation technique discussed, to the best of our knowledge, was not derived explicitly before. The experimental setup which we used for a verification was inspired by the Mach-Zehnder quantum eraser setup reviewed in the paper by Schneider et al. (2001) [4].

[1*]Teledyne FLIR LLC, 1049 Camino Dos Rios Thousand Oaks, CA 91360, USA
✉ artem.kryvobok@teledyneflir.com
[2]Teledyne FLIR LLC, 1049 Camino Dos Rios Thousand Oaks, CA 91360, USA

On behalf of all authors, the corresponding author states that there is no conflict of interest.
All data generated or analysed during this study are included in this published article (and its supplementary information files).

This is an elegant approach for an interference type experiment involving investigation of the polarization states of light. It is also a valid approach to emulate single-photon interference statistics. By splitting the classical light into two parts the Mach-Zehnder introduces a definite phase relation between the parts at a recombining beamsplitter. Additionally, from purely formal position it is a valid method due perfect interchangeability of displacement and Fock-state operators having no effect on the matrix components. This, in turn, allows investigation of the interference statistics equivalently for classical and non-classical light. Hence, the developed 4-dimensional photonic Hadamard-like transformation is applicable to single-photon based quantum computing operations and simulations.

## II. Theoretical setup

The beamsplitter operation is modeled by a matrix acting on the input state and transforming it in this way to the output state by matrix multiplication. The states are represented in a 4-dimensional vector space comprising of two spatial modes – input/output ports of the beamsplitter and two transverse polarization modes of the beam, which we indicate with subscript $v$ and $h$. Hence, let's start derivation of our non-polarizing beamsplitter as a 4x4 matrix of the form (1) [5].

$$\begin{pmatrix} t_v & r_v & 0 & 0 \\ r_v & t_v & 0 & 0 \\ 0 & 0 & t_h & r_h \\ 0 & 0 & r_h & t_h \end{pmatrix} \tag{1}$$

The constituting matrix coefficients $t_v, t_h, r_v, r_h$ are amplitudes for transmission of vertical and horizontal polarization and reflection of vertical and horizontal polarization accordingly. For a balanced (50/50) beamsplitter we set the magnitudes of the amplitudes equal: $|t_v| = |t_h| = |r_v| = |r_h|$. Since there is no birefringence assumed within a beamsplitter we also set zero relative phase between $t_v$ and $t_h$, so for a simplicity: $t_v = t_h = t$. Additionally, any phase shift related to the propagation velocity within the beamsplitter is equivalent for all the coefficients. It is therefore convenient to treat the phase of $t$ as the phase reference for $r_v$ and $r_h$, so we let $t > 0, t\epsilon\mathbb{R}$ with respect to the reflection coeffiecients. These definitions form our pre-experimental ansatz. Anticipating the experiment results suggested reflection amplitudes for the same polarization state of different spatial modes are not necessarily equal. Hence, as a part of the post-experimental ansatz we set $r_v$ and $\widetilde{r_v}$ to be the reflection amplitudes of different ports for vertical polarization and $r_h$ and $\tilde{r}_h$ – the reflection amplitudes of different ports for horizontal polarization accordingly. Further, the $r_v$, $\tilde{r}_v$, $r_h$ and $\tilde{r}_h$ are understood to be complex numbers. Thus, in amplitude – phase factor representation [15] the coefficients are defined as: $r_v = |r_v|\exp(i\varphi_{r_v})$, $\tilde{r}_v = |\tilde{r}_v|\exp(i\varphi_{\tilde{r}_v})$, $r_h = |r_h|\exp(i\varphi_{r_h})$, $\tilde{r}_h = |\tilde{r}_h|\exp(i\varphi_{\tilde{r}_h})$ and $t = |t|\exp(i\varphi_t)$, $|t_v| = |t_h| = |r_v| = |r_h| = |\tilde{r}_v| = |\tilde{r}_h|$. Wherein we set $\varphi_t = 0$ treating the transmission phase as a reference relative to phases of reflection coefficients. Consequently, the resulting matrix ($B$) is of the following form:

$$B = \begin{pmatrix} t & r_v & 0 & 0 \\ \tilde{r}_v & t & 0 & 0 \\ 0 & 0 & t & r_h \\ 0 & 0 & \tilde{r}_h & t \end{pmatrix} \tag{2}$$

A physical representation of input and output ports of a beamsplitter is shown on Fig. 1.

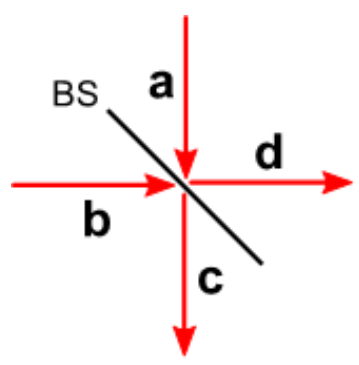

Fig. 1. Input and output ports in a beamsplitter.

We define the input state having input spatial ports ***a*** and ***b*** and transverse modes ***v*** and ***h*** with normalized amplitudes $a$, $b$, $v$ and $h$ accordingly:

$$|\psi_{in}\rangle \rightarrow (a|a\rangle + b|b\rangle) \otimes (v|v\rangle + h|h\rangle) \rightarrow$$

$$a_v|a_v\rangle + b_v|b_v\rangle + a_h|a_h\rangle + b_h|b_h\rangle = \begin{pmatrix} a_v \\ b_v \\ a_h \\ b_h \end{pmatrix} \tag{3}$$

where observing $\langle\psi_{in}|\psi_{in}\rangle = 1$, $a_v$ and $a_h$ are normalized amplitudes in modes $\boldsymbol{a_v}$ and $\boldsymbol{a_h}$ for vertical and horizontal polarizations in the input port a. Further, $b_v$ and $b_h$ are normalized amplitudes in modes $\boldsymbol{b_v}$ and $\boldsymbol{b_h}$ for vertical and horizontal polarizations in the input port b. The output state is then obtained by $B$ acting on the input state:

$$|\psi_{out}\rangle = B|\psi_{in}\rangle = \begin{pmatrix} c_v \\ d_v \\ c_h \\ d_h \end{pmatrix} = \begin{pmatrix} ta_v + r_v b_v \\ \tilde{r}_v a_v + tb_v \\ ta_h + r_h b_h \\ \tilde{r}_h a_h + tb_h \end{pmatrix} \tag{4}$$

where, according to the convention (Fig.1), the amplitudes $c_v, c_h, d_v$ and $d_h$ of the corresponding modes $\boldsymbol{c_v}, \boldsymbol{c_h}, \boldsymbol{d_v}$ and $\boldsymbol{d_h}$ in the output ports c and d are as well normalized to satisfy $\langle\psi_{out}|\psi_{out}\rangle = 1$. Now, following the principle of conservation of energy, a lossless beamsplitter should perform a unitary transformation requiring:

$$|\psi_{in}|^2 = |\psi_{out}|^2 \tag{5}$$

The expressions (4) and (5) lead to explicit unitarity condition for a beamsplitter:

$$|\psi_{out}|^2 = \langle\psi_{in}|B^{\dagger}B|\psi_{in}\rangle \tag{6}$$

where Eq. (6) can only be satisfied if:

$$B^{\dagger}B = I_4 \tag{7}$$

with $I_4$ being a unit 4x4 matrix. The equation (7) leads to the following set of equations (8,a-h) which are sometimes referred to as reciprocity relations [6]:

$$|t|^2 + |\tilde{r}_v|^2 = 1 \qquad (8,a)$$

$$|t|^2 + |r_v|^2 = 1 \qquad (8,b)$$

$$|t|^2 + |\tilde{r}_h|^2 = 1 \qquad (8,c)$$

$$|t|^2 + |r_h|^2 = 1 \qquad (8,d)$$

$$\text{Since: } t > 0, t\epsilon\mathbb{R} \Rightarrow |t_v| = |t_h| = t = \frac{\mathbf{1}}{\sqrt{\mathbf{2}}} \Rightarrow t^* = t;$$

$$t = |r_v| = |r_h| \Rightarrow$$

$$r_v + \tilde{r}_v^{\,*} = 0 \Rightarrow \exp(i\varphi_{r_v}) = \exp(\pm i\pi - i\varphi_{\tilde{r}_v}) \qquad (8,e)$$

$$r_v^* + \tilde{r}_v = 0 \Rightarrow \exp(-i\varphi_{r_v}) = \exp(\pm i\pi + i\varphi_{\tilde{r}_v}) \qquad (8,f)$$

$$r_h + \tilde{r}_h^{\,*} = 0 \Rightarrow \exp(i\varphi_{r_h}) = \exp(\pm i\pi - i\varphi_{\tilde{r}_h}) \qquad (8,g)$$

$$r_h^* + \tilde{r}_h = 0 \Rightarrow \exp(-i\varphi_{r_h}) = \exp(\pm i\pi + i\varphi_{\tilde{r}_h}) \qquad (8,h)$$

Recapping on the amplitudes conditions set out earlier: $t > 0, t\epsilon\mathbb{R} \Rightarrow |t_v| = |t_h| = t$, $t = |r_v| = |r_h|$ and $r_v, \tilde{r}_v, r_h, \tilde{r}_h \epsilon \mathbb{Z}$. Thus, we obtain the following set of possible solutions satisfying (8,a-h):

$$r_v = \tilde{r}_v = \frac{\pm i}{\sqrt{2}} \; and \; r_h = \tilde{r}_h = \frac{\pm i}{\sqrt{2}} \qquad (8,j)$$

$$r_v = \tilde{r}_v = \frac{\pm i}{\sqrt{2}} \; and \; r_h = \tilde{r}_h = \frac{\mp i}{\sqrt{2}} \qquad (8,i)$$

$$\varphi_{r_v} = \varphi_{\tilde{r}_v} = \frac{\pm\pi}{2} \; and \; \varphi_{r_h} = \varphi_{\tilde{r}_h} = \frac{\pm\pi}{2} \qquad (8,k)$$

$$\varphi_{r_v} = \varphi_{\tilde{r}_v} = \frac{\pm\pi}{2} \; and \; \varphi_{r_h} = \varphi_{\tilde{r}_h} = \frac{\mp\pi}{2} \qquad (8,l)$$

$$r_v = \frac{\pm 1}{\sqrt{2}} \; then \; \tilde{r}_v = \frac{\mp 1}{\sqrt{2}} \; and \; r_h = \frac{\pm 1}{\sqrt{2}} \; and \; \tilde{r}_h = \frac{\mp 1}{\sqrt{2}} \qquad (8,m)$$

$$r_v = \frac{\pm 1}{\sqrt{2}} \; then \; \tilde{r}_v = \frac{\mp 1}{\sqrt{2}} \; and \; r_h = \frac{\mp 1}{\sqrt{2}} \; and \; \tilde{r}_h = \frac{\pm 1}{\sqrt{2}} \qquad (8,n)$$

$$\begin{matrix} \varphi_{r_v} = 0 \; then \; \varphi_{\tilde{r}_v} = \pi \; and \; \varphi_{r_h} = 0 \; and \; \varphi_{\tilde{r}_h} = \pi \\ \varphi_{r_v} = \pi \; then \; \varphi_{\tilde{r}_v} = 0 \; and \; \varphi_{r_h} = \pi \; and \; \varphi_{\tilde{r}_h} = 0 \end{matrix} \qquad (8,o)$$

$$\begin{matrix} \varphi_{r_v} = 0 \; then \; \varphi_{\tilde{r}_v} = \pi \; and \; \varphi_{r_h} = \pi \; and \; \varphi_{\tilde{r}_h} = 0 \\ \varphi_{r_v} = \pi \; then \; \varphi_{\tilde{r}_v} = 0 \; and \; \varphi_{r_h} = 0 \; and \; \varphi_{\tilde{r}_h} = \pi \end{matrix} \qquad (8,p)$$

Hence, if solely condition (7) is considered the matrix (2) can be regarded as two 2x2 zero matrices and two 2x2 independent matrices acting separately on different transverse modes. However, as it was mentioned earlier the experimental evidence showed a correlation between $\tilde{r}_v$ and $\tilde{r}_h$, $r_v$ and $r_h$ as well as between $\tilde{r}_v$ and $r_v$, $\tilde{r}_h$ and $r_h$. I.e. the two 2x2 matrices acting on separate transverse modes are not independent. Namely, it was verified that only solutions (8,n) and (8,p) accordingly agree with the experimental results.

We found that a rigid theoretical formulation of these solutions can be provided either by the Fresnel equations [19]

or by the framework of quantum optics, i.e. when considering the second quantization of the electromagnetic fields. For a full quantum mechanical treatment of the subject we opted for the latter approach. Namely, we replace the field amplitudes in (3) and (4) with the Fock state photon operators and impose the canonical commutation relations onto them as per general case:

$$\left[\hat{a}_s, \hat{a}_{s'}^\dagger\right] = \delta_{ss'} \tag{9}$$

where $\hat{a}_s$ is a photon annihilation operator in mode $s$ and $\hat{a}_{s'}^\dagger$ is a photon creation operator in mode $s'$. These commutation relations were demonstrated to hold for the two output modes in a balanced 2-dimensional beamsplitter [20]. We, in turn, considered this approach in a balanced 4-dimensional beamsplitter. First, let us switch the output modes: $\boldsymbol{c_v}, \boldsymbol{c_h}, \boldsymbol{d_v}$ and $\boldsymbol{d_h}$ into a Fock space representation. Taking into account that the fields are tensor products of spatial and transverse vectors, i.e.: $|\psi_{out}\rangle \rightarrow (|c\rangle + |d\rangle) \otimes (|v\rangle + |h\rangle)$, we identify photons in each separate mode:

$$\boldsymbol{c_v} \rightarrow |1_c, 1_v\rangle = \hat{a}_c^\dagger \hat{a}_v^\dagger |0\rangle \tag{10,a}$$

$$\boldsymbol{d_v} \rightarrow |1_d, 1_v\rangle = \hat{a}_d^\dagger \hat{a}_v^\dagger |0\rangle \tag{10,b}$$

$$\boldsymbol{c_h} \rightarrow |1_c, 1_h\rangle = \hat{a}_c^\dagger \hat{a}_h^\dagger |0\rangle \tag{10,c}$$

$$\boldsymbol{d_h} \rightarrow |1_d, 1_h\rangle = \hat{a}_d^\dagger \hat{a}_h^\dagger |0\rangle \tag{10,d}$$

where $\hat{a}_c^\dagger$, $\hat{a}_d^\dagger$, $\hat{a}_v^\dagger$ and $\hat{a}_h^\dagger$ are photon creation operators in spatial modes $\boldsymbol{c}$, $\boldsymbol{d}$ and in transverse modes $\boldsymbol{v}$, $\boldsymbol{h}$ accordingly and $|0\rangle$ is the vacuum state. In the 2-dimensional case the commutation relation given in [20] concerned two spatial modes, while with the tensor product we group the commutator per particular mode. According to the order given in the output state (4) we have the following commutation relation to hold:

$$\left[\hat{a}_c\hat{a}_v + \hat{a}_d\hat{a}_v, \hat{a}_c^\dagger\hat{a}_h^\dagger + \hat{a}_d^\dagger\hat{a}_h^\dagger\right] = 0 \Rightarrow$$

$$\left[\hat{a}_c\hat{a}_v, \hat{a}_c^\dagger\hat{a}_h^\dagger\right] + \left[\hat{a}_c\hat{a}_v, \hat{a}_d^\dagger\hat{a}_h^\dagger\right] + \left[\hat{a}_d\hat{a}_v, \hat{a}_c^\dagger\hat{a}_h^\dagger\right] +$$

$$\left[\hat{a}_d\hat{a}_v, \hat{a}_d^\dagger\hat{a}_h^\dagger\right] = 0 \tag{11,a}$$

where $\hat{a}_c$, $\hat{a}_d$ and $\hat{a}_v$ photon annihilation operators in spatial modes $\boldsymbol{c}$, $\boldsymbol{d}$ and in transverse mode $\boldsymbol{v}$ accordingly. Now expanding the output modes in (11,a) into coupled input modes:

$$\left[\hat{a}_c\hat{a}_v, \hat{a}_c^\dagger\hat{a}_h^\dagger\right] = \left[t\hat{a}_a\hat{a}_v + r_v\hat{a}_b\hat{a}_v, t^*\hat{a}_a^\dagger\hat{a}_h^\dagger + r_h^*\hat{a}_b^\dagger\hat{a}_h^\dagger\right] \tag{11,b}$$

$$\left[\hat{a}_c\hat{a}_v, \hat{a}_d^\dagger\hat{a}_h^\dagger\right] = \left[t\hat{a}_a\hat{a}_v + r_v\hat{a}_b\hat{a}_v, \widetilde{r_h}^*\hat{a}_a^\dagger\hat{a}_h^\dagger + t^*\hat{a}_b^\dagger\hat{a}_h^\dagger\right] \tag{11,c}$$

$$\left[\hat{a}_d\hat{a}_v, \hat{a}_c^\dagger\hat{a}_h^\dagger\right] = \left[\widetilde{r_v}\hat{a}_a\hat{a}_v + t\hat{a}_b\hat{a}_v, t^*\hat{a}_a^\dagger\hat{a}_h^\dagger + r_h^*\hat{a}_b^\dagger\hat{a}_h^\dagger\right] \tag{11,d}$$

$$\left[\hat{a}_d\hat{a}_v, \hat{a}_d^\dagger\hat{a}_h^\dagger\right] = \left[\widetilde{r_v}\hat{a}_a\hat{a}_v + t\hat{a}_b\hat{a}_v, \widetilde{r_h}^*\hat{a}_a^\dagger\hat{a}_h^\dagger + t^*\hat{a}_b^\dagger\hat{a}_h^\dagger\right] \tag{11,e}$$

where $\hat{a}_a$, $\hat{a}_b$, are photon annihilation operators in input modes $\boldsymbol{a}, \boldsymbol{b}$ and $\hat{a}_a^\dagger$, $\hat{a}_b^\dagger$, are photon creation operators in input modes $\boldsymbol{a}, \boldsymbol{b}$ accordingly. Making use of (9) to solve (11,a) we obtain the following identities (explicit derivation of (11,a) is given in Appendix I):

$$|t|^2 + r_v r_h^* = 0 \tag{11,f}$$

$$t\widetilde{r_h}^* + r_v t^* + \widetilde{r_v} t^* + t r_h^* = 0 \tag{11,g}$$

$$\widetilde{r_v}\widetilde{r_h}^* + |t|^2 = 0 \tag{11,h}$$

From the above equations we can effectively obtain the necessary correlations between all the amplitudes $\widetilde{r_v}$ and $\widetilde{r_h}$, $r_v$ and $r_h$. Considering: $t > 0, t\epsilon\mathbb{R}$ (11,f-h) and (8,a-h) are satisfied if and only if:

$$if\ r_h = \frac{-1}{\sqrt{2}}\ then\ \widetilde{r_h} = \frac{1}{\sqrt{2}}\ and\ r_v = \frac{1}{\sqrt{2}}\ and\ \widetilde{r_v} = \frac{-1}{\sqrt{2}};$$

or

$$if\ r_h = \frac{1}{\sqrt{2}}\ then\ \widetilde{r_h} = \frac{-1}{\sqrt{2}}\ and\ r_v = \frac{-1}{\sqrt{2}}\ and\ \widetilde{r_v} = \frac{1}{\sqrt{2}}.$$

These solutions require the relative phases of the reflection amplitudes to strictly satisfy:

$$if\ \varphi_{r_v} = \pi\ then\ \varphi_{\widetilde{r_v}} = 0\ and\ \varphi_{r_h} = 0\ and\ \varphi_{\widetilde{r_h}} = \pi;$$

or

$$if\ \varphi_{r_v} = 0\ then\ \varphi_{\widetilde{r_v}} = \pi\ and\ \varphi_{r_h} = \pi\ and\ \varphi_{\widetilde{r_h}} = 0.$$

Hence, the sought matrix can take either of the following forms:

$$B = \frac{1}{\sqrt{2}}\begin{pmatrix} 1 & -1 & 0 & 0 \\ 1 & 1 & 0 & 0 \\ 0 & 0 & 1 & 1 \\ 0 & 0 & -1 & 1 \end{pmatrix} \tag{12,a}$$

$$B = \frac{1}{\sqrt{2}}\begin{pmatrix} 1 & 1 & 0 & 0 \\ -1 & 1 & 0 & 0 \\ 0 & 0 & 1 & -1 \\ 0 & 0 & 1 & 1 \end{pmatrix} \tag{12,b}$$

These expressions agree with our experimental results and the proposed approach of applying the commutation relations to the output modes of a 4-dimensional beamsplitter is, to the best of our knowledge, novel. One should also note that the relations provide full set of correct solutions needless to combine with (8,a-h). Therefore, the commutations technique proposed can be regarded as a sufficient condition on its own.

## III. Experimental setup and results

In order to test the derived matrix expressions the input state of a beamsplitter should be prepared to have two spatial modes with definite initial phase as well as various combinations of polarization modes. Generally speaking, any two spatial input modes with correlated relative phase and controlled polarization modes satisfy the required input state. From a practical point of view, however, the easiest solution to both prepare the input state accordingly and test a beamsplitter action can be accomplished by the Mach-Zehnder interferometer (MZ) with a single coherent input. The part of such MZ just before a recombining beamsplitter provides two spatial modes with correlated relative phase and controlled polarization modes effectively prepares the input state as necessary while a recombing beamsplitter plays a role of a test beamsplitter. The details of our MZ experimental setup are shown on Fig. 2. In this setup we investigate the correlations of intensity oscillations of output arms from the ports 4 and 5 of MZ. The oscillations allow fully defining the output state and hence determine unambiguously an action of a beamsplitter upon the input state, i.e. the exact phase and amplitude relations within the beamsplitter matrix.

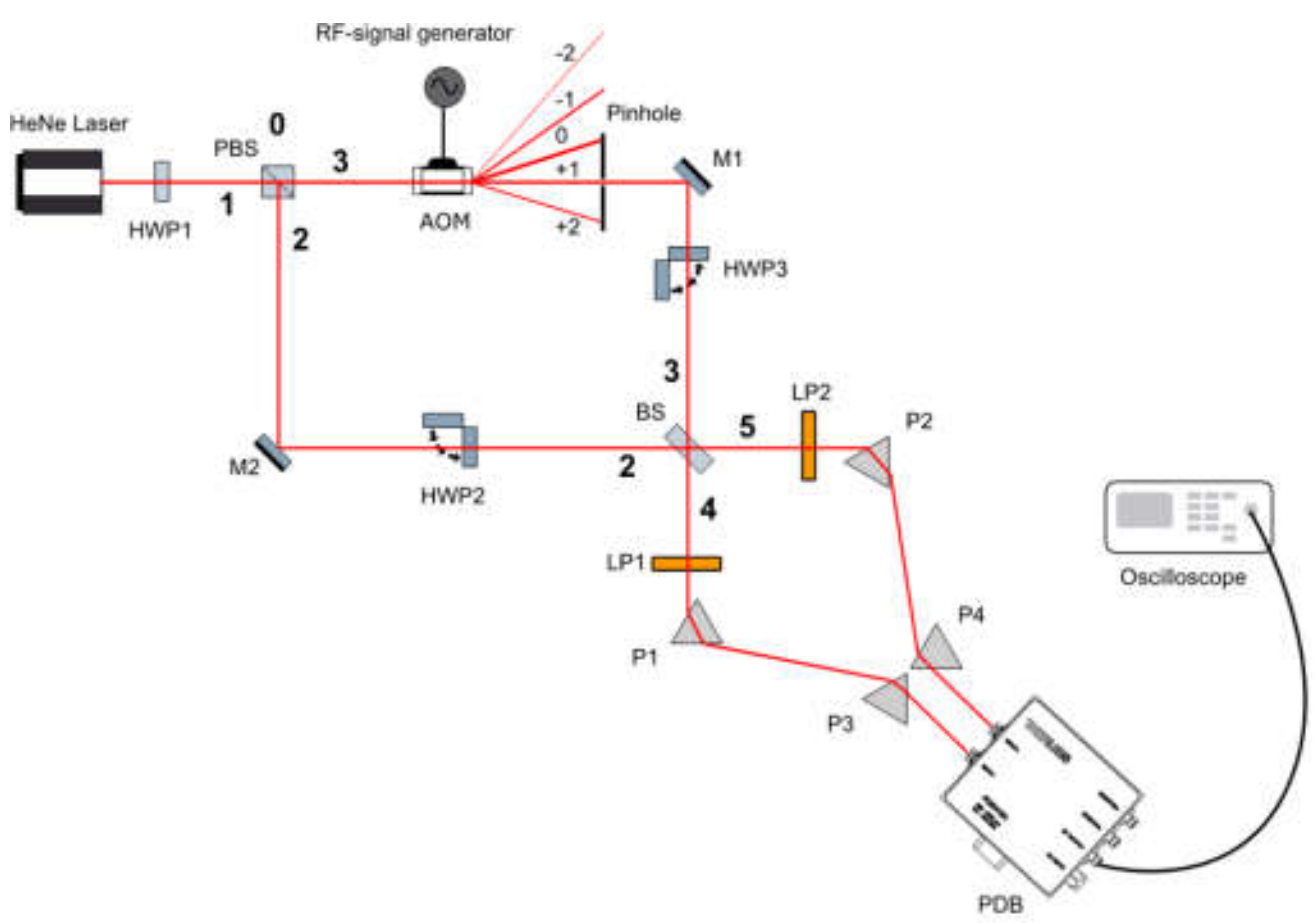

Fig. 2. Experimental setup testing the ansatz of beamsplitter matrix.

As a single coherent input we used 5 mW HeNe laser at 632.8 nm. The laser was specified to have linearly polarized, single transverse mode Gaussian profile beam with linewidth of 1.8 MHz. This ensured a sufficient coherence length of several meters. Past the HeNe we inserted the first half-wave plate (HWP1) so to ease the control of linear polarization of the beam before MZ. The beam enters MZ in the polarizing beamsplitter (PBS) which transmits horizontally polarized component (port 3) of the beam and reflects the vertically polarized component (port 2). The PBS features an extinction ratio of 1:1000 according to specifications, so for our laser we can expect a substantial polarization purity in our measurements. In order to introduce a regular phase oscillation we inserted an acousto-optic modulator (AOM) in horizontally polarized arm of MZ. This imposed continuous phase oscillation in RF-band, so lower frequency oscillations caused by ambient lab vibrations and thermal changes induce slowly-varying envelope and can be neglected in our measurements. The choice of polarization in this case was arbitrary. We drove the AOM with the RF-signal generator tuned to a fixed frequency of about 107 MHz with 24 V peak-to-peak voltage. This provided the required power of acoustic beam inside the AOM to achieve high first-order diffraction efficiency. The particular optical beam and the AOM model parameters lead to operating in the Raman-Nath acousto-optic regime [7] producing several output diffraction orders. The diffracted beams are shifted in frequency equivalent to the diffraction order, i.e. 107 MHz multiple of: -2, -1, 0, +1, +2. For our measurements we chose the +1 order, i.e. up-shifted by about 107 MHz respecting the source beam. Since intensity fluctuations investigated are caused by beat the -1 order could have been equally chosen. We placed a pinhole with variable iris in the selected diffracted beam to isolate it for further alignment. The pinhole opening was slightly smaller than resulting beam diameter. Thus, due to Gaussian intensity distribution of wave front we could ensure bullseye interference alignment when beat amplitude was at highest amplitude. The resulting arms of MZ were directed by mirrors M1 and M2 to recombine at Thorlabs BS016 beamsplitter (BS).

It is important to note that while BS016 has ±10% polarization selectivity and is ±3% lossy [18] at 632.8nm we are going to treat it as a balanced non-polarizing beamsplitter when estimating the agreement with the theory. This is justified by the fact that if we include the arbitrary transmission and reflection coefficients for s- and p-pol modes it is then not a non-polarizing beamsplitter, i.e. conflicting with the theoretical concept we discuss.

Each inner arm of MZ could have been switched to different polarization state by insertion of half-wave plates HWP2 and HWP3. The HWP2 and HWP3 had their fast axis (FA) set at $\frac{\pi}{4}$ rad. Thus upon insertion of HPW2 or HWP3 we could obtain: both vertical; both horizontal states; and swapping the states by inserting both waveplates before the recombination in BS. Further, past BS we set two linear polarizers: LP1 and LP2, i.e. in ports 4 and 5 accordingly. The relative orientations of the LP1 and LP2 axes played role in measuring phase amplitudes in our experiment. Thus, when taking measurements with two arms polarized orthogonally by setting the linear polarizers axes at $\frac{\pi}{4}$ rad or $-\frac{\pi}{4}$ rad we could superpose diagonal components of orthogonal states leading to interference and producing temporal intensity fluctuations with phase oscillations from the AOM. When HWP2 or HWP3 was inserted in either arm shifting them to the same polarization state LP1 and LP2 were set both accordingly to either vertical or horizontal position, so to achieve maximum transmission level, i.e. both at either 0 or $\frac{\pi}{2}$ rad. Hence, the above discussed HWP2, HWP3, LP1 and LP2 configurations produced beat on output arms 4 and 5. The correlations of beat intensities were observed by sending output beams individually into two P-i-N type photodetectors (PD). The output arms were steered at normal incidence to PD's inputs using arrangements of equilateral dispersive prisms P1, P2 and anamorphic prisms P3 and P4 as shown on Fig.2. The prisms were mounted onto kinematic mounts with azimuthal and polar degrees of freedom so beams could be conveniently aligned with PD's surfaces. The equilateral prisms provided a necessary sharp deviation angle which helped placing PD's in a compact manner yet the prisms suffered some polarization selectivity. As we study the action of unpolarized BS this adds a certain degree of complexity. According to the manufacturer specifications about 27% of transverse electric (TE) mode intensity [8] is reflected from the surface of the prism, so we correlated vertical component of a field amplitude to a factor of $\sqrt{0.73}$ per surface in our calculations. The second bend step was accomplished with anamorphic prisms featuring sufficient deviation and not having any polarization selectivity. The two PD's were housed inside balanced amplified photodetector (PBD). We used the Thorlabs fixed gain PDB410A model PBD with 100 MHz bandwidth suitable for modulation frequency of the AOM. It is important to emphasize that using a PBD detection is an essential part of the experiment in terms of providing an unambiguous phase correlation of output beams. This is due to an accurate spatial match and match of response time of both PD's. The PBD had two PD's photocurrent outputs coupled into a single RF output. One of the PD's operated in reverse bias ("input -") while the other in forward bias ("input +"). Hence, the PD's generated photocurrent in opposite directions relative to each other. In such a way when two beams of the same intensity are shone upon the PD's their individual photocurrents cancel each other out and the resulting current output is 0. Conversely, when there is an intensity difference the current level shifts to either negative or positive side depending on the PD bias direction. The PBD output was connected to a digital oscilloscope with 200 MHz spectral bandwidth, which fully accommodated our modulation frequency. The experimental results were, hence, obtained in a form of oscillograms.

The measurements were performed for all possible polarization configurations, to which we will refer as six test configurations throughout this paper, in order to comprehensively test the ansatz on BS matrix. The six configurations can be summarized in the following Table I.:

TABLE I.
SUMMARY OF TEST CONFIGURATIONS

| Test configuration | Waveplates | | Resulting state before the recombination in beamsplitter | | Position of linear polarizers, rad | |
|---|---|---|---|---|---|---|
| | HWP2 | HWP3 | Arm 2 | Arm 3 | LP1 | LP2 |
| 1 | in | out | Horizontal | Horizontal | $\frac{\pi}{2}$ | $\frac{\pi}{2}$ |
| 2 | out | in | Vertical | Vertical | 0 | 0 |
| 3 | out | out | Vertical | Horizontal | $\frac{\pi}{4}$ | $\frac{\pi}{4}$ |
| 4 | out | out | Vertical | Horizontal | $\frac{\pi}{4}$ | $-\frac{\pi}{4}$ |
| 5 | in | in | Horizontal | Vertical | $\frac{\pi}{4}$ | $\frac{\pi}{4}$ |
| 6 | in | in | Horizontal | Vertical | $\frac{\pi}{4}$ | $-\frac{\pi}{4}$ |

The oscillograms produced for the six test configurations are provided on Fig.3.1-6 below:

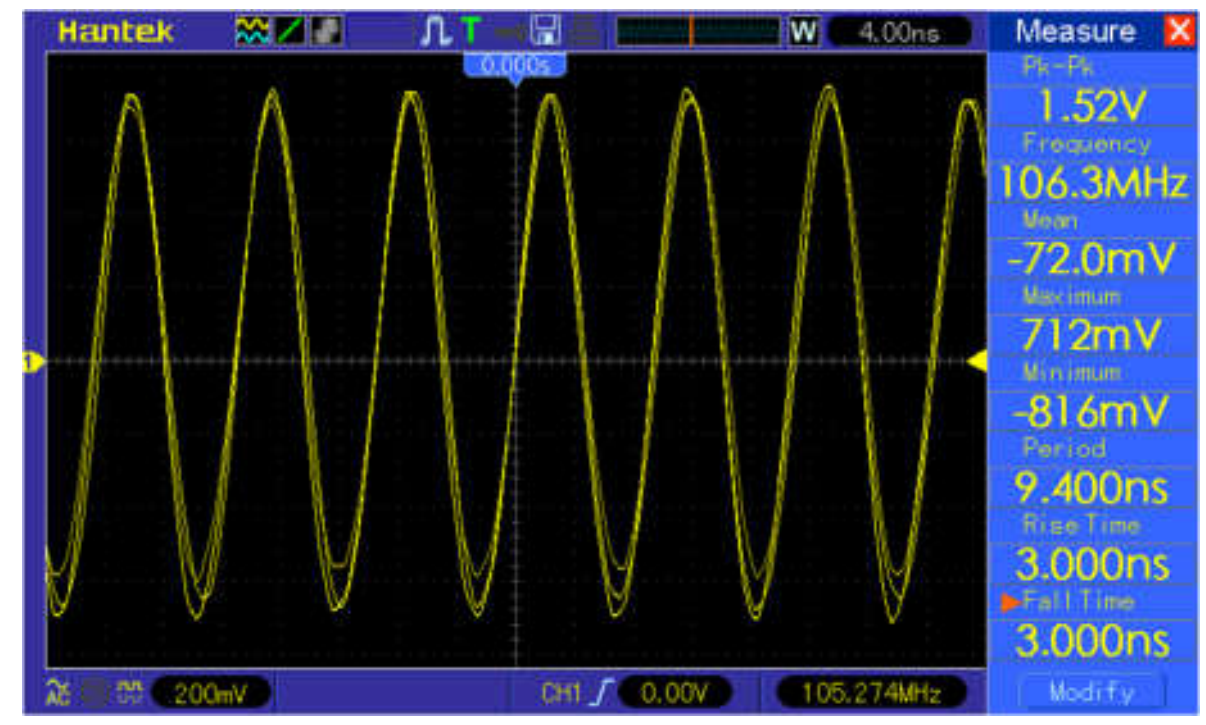

Fig. 3.1. Oscillation amplitude with both arms recombined in the horizontal state with LP1 and LP2 set at $\frac{\pi}{2}$ rad.

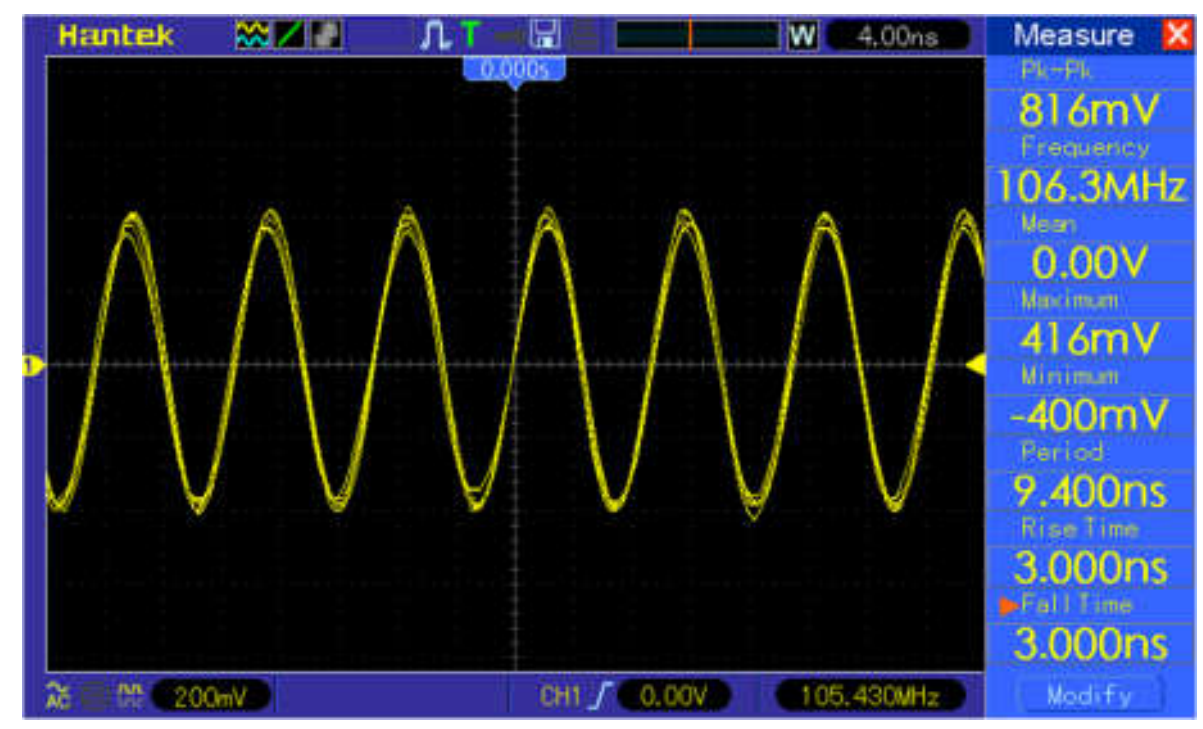

Fig. 3.2. Oscillation amplitude with both arms recombined in the vertical state with LP1 and LP2 set at 0 rad.

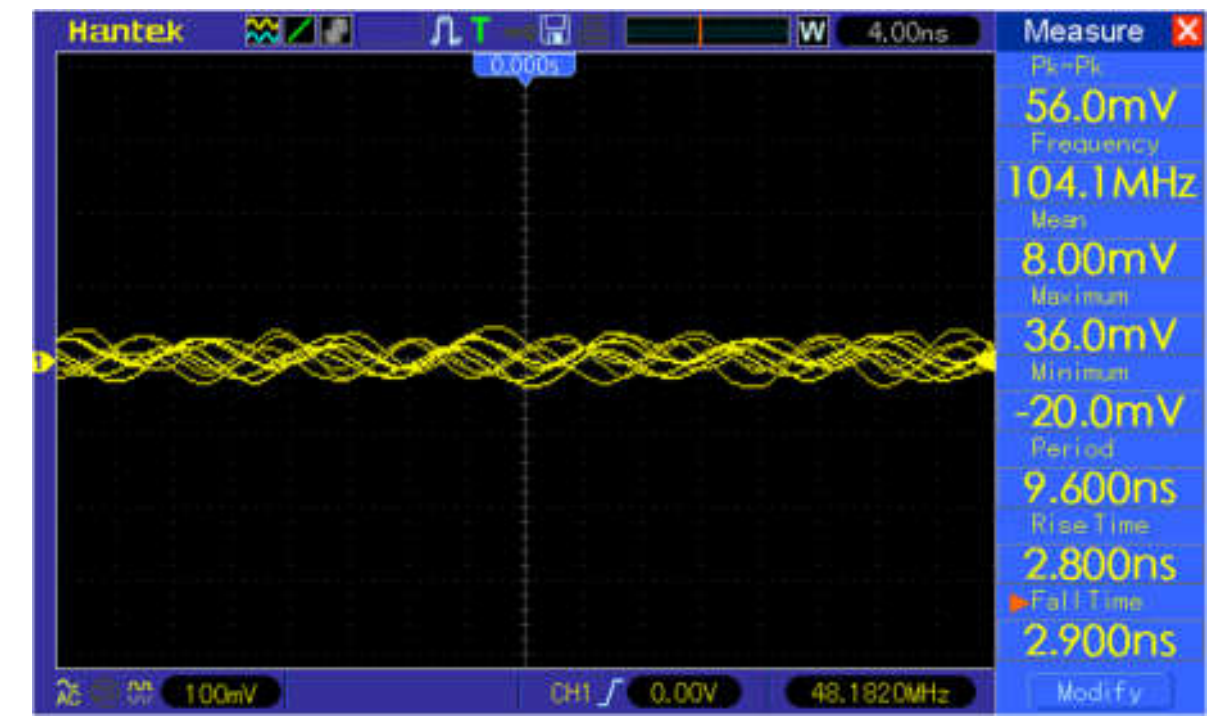

Fig. 3.3. Oscillation amplitude produced by recombining vertically polarized arm 2 and horizontally polarized arm 3 with LP1 and LP2 both set symmetrically at $\frac{\pi}{4}$ rad.

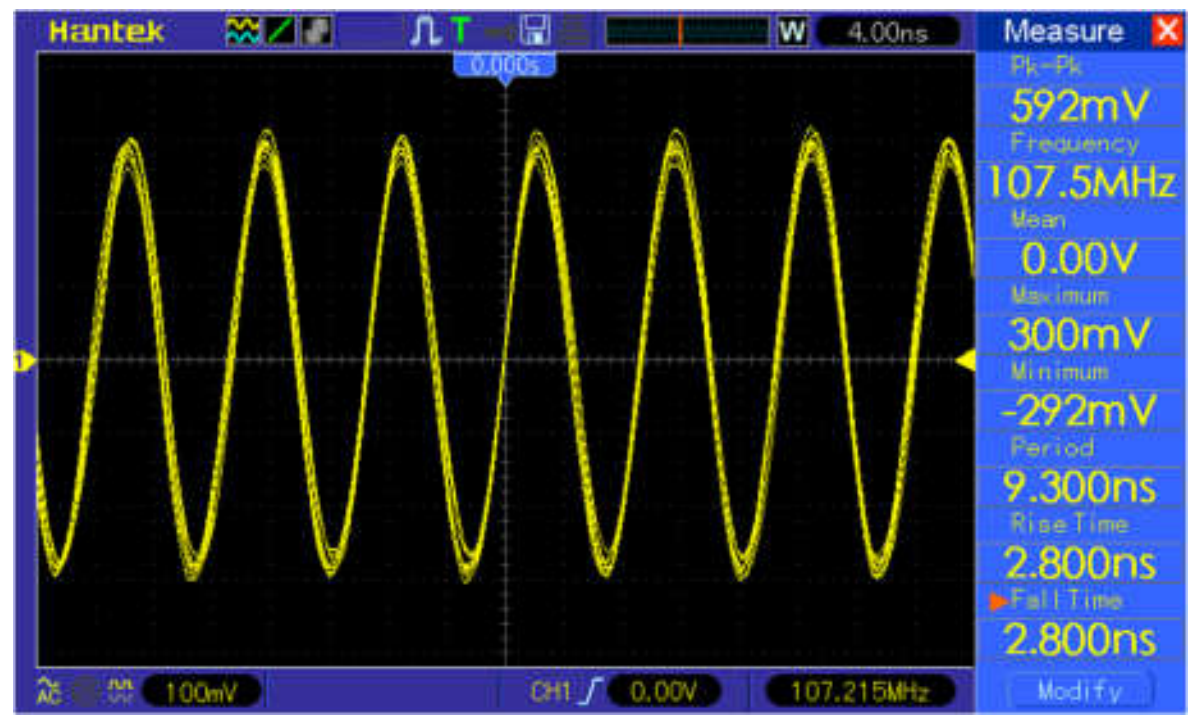

Fig. 3.4. Oscillation amplitude produced by recombining vertically polarized arm 2 and horizontally polarized arm 3 with LP1 and LP2 set anti-symmetrically, i.e. LP1 at $\frac{\pi}{4}$ rad and LP2 at $-\frac{\pi}{4}$ rad.

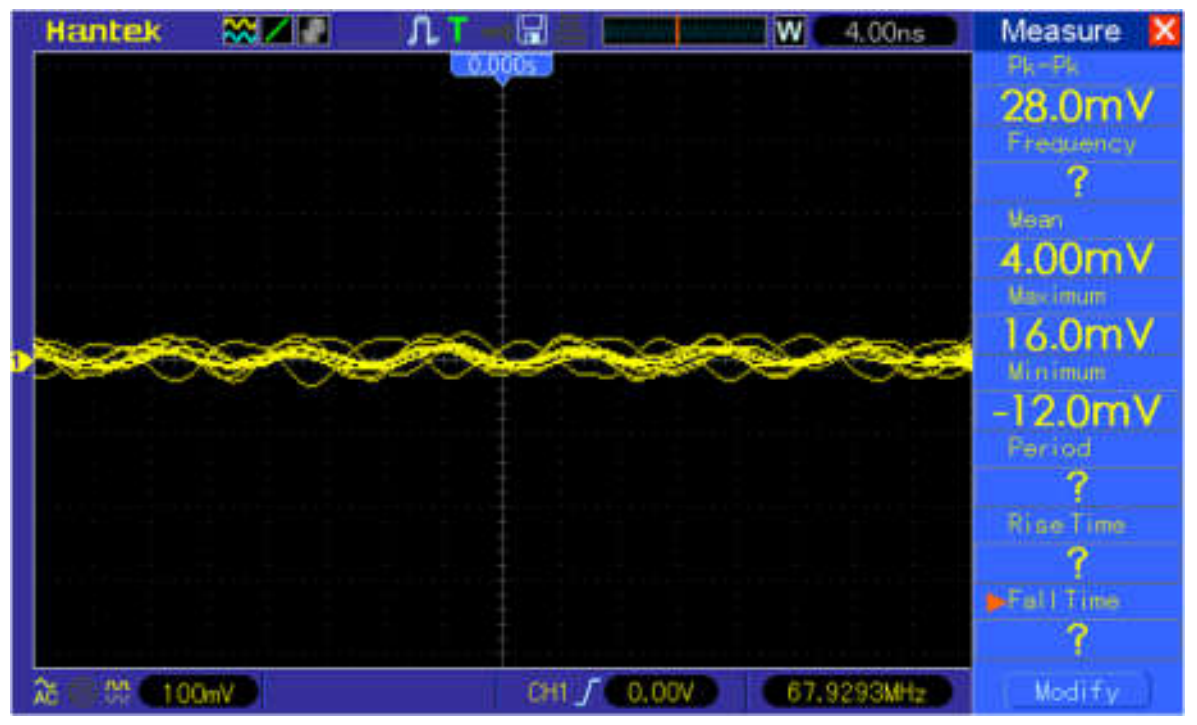

Fig. 3.5. Oscillation amplitude produced by recombining horizontally polarized arm 2 and vertically polarized arm 3 with LP1 and LP2 set symmetrically, i.e. LP1 at $\frac{\pi}{4}$ rad and LP2 at $\frac{\pi}{4}$ rad.

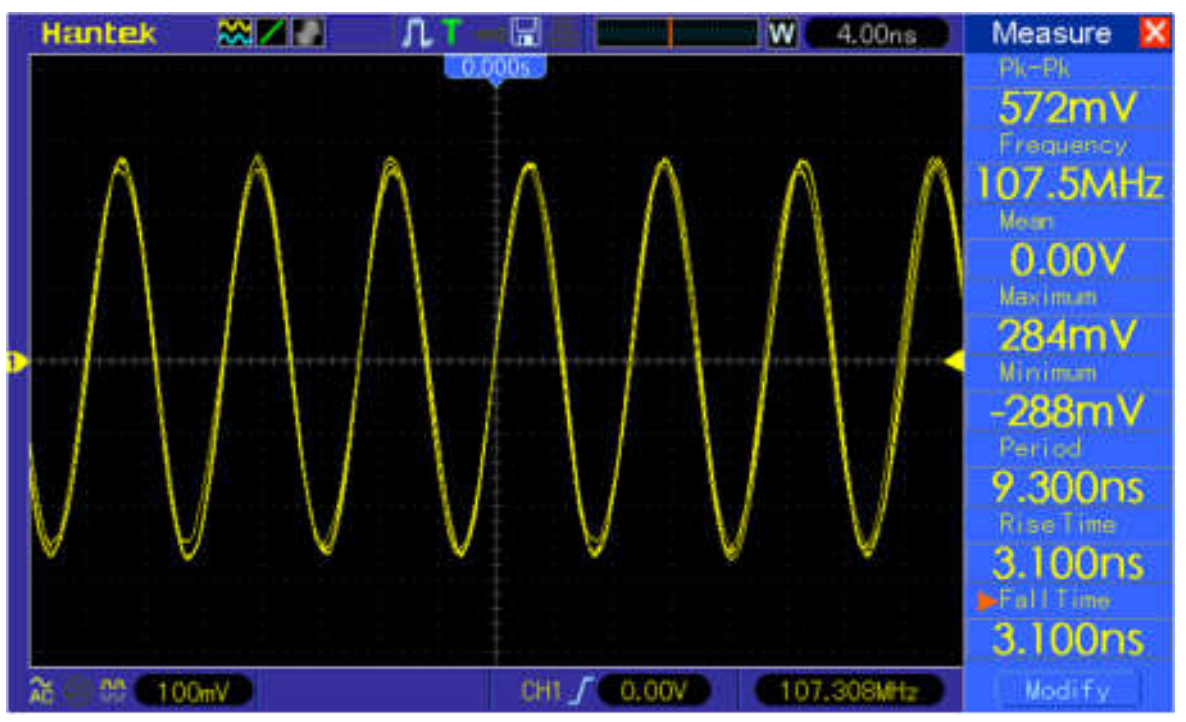

Fig. 3.6. Oscillation amplitude produced by recombining horizontally polarized arm 2 and vertically polarized arm 3 with LP1 and LP2 set anti-symmetrically, i.e. LP1 at $\frac{\pi}{4}$ rad and LP2 at $-\frac{\pi}{4}$ rad.

## IV. THEORETICAL AGREEMENT

In order to match the theoretical predictions with the format of the experimental results we plotted the temporary interference intensity for each configuration. The temporal interference is given as a voltage oscillation amplitude per full phase cycle of the AOM. The interference intensity, in turn, was derived from the quantum mechanical evolution of the state vector using formalism described in eq. 4 and 12. The detailed description of the derivation steps is given in Appendix II including the reference to the factor 0.73 present in calculations (as explained in the experimental part due to the polarization selectivity of certain optical instruments). We will use the expression (12,a) as a test BS matrix for the purposes of showing worked example here. However, both definitions (12,a-b) produce the same phase correlations between observables.

For the first test configuration the output state is:

$$|\psi_{out}^{I}\rangle \rightarrow \frac{1}{2}\begin{pmatrix} 0 \\ 0 \\ (1+e^{-i\varphi})\widehat{D}_{4,h}(\alpha_{4,h}) \\ (e^{-i\varphi}-1)\widehat{D}_{5,h}(\alpha_{5,h}) \end{pmatrix}|0\rangle \qquad (25)$$

The beamsplitter matrices are not Hermitian so we find intensities as probability density per port rather than an expectation value. Hence, the normalized intensity in port 4 oscillates as: $I_4^I(\varphi) = \left|\frac{1}{2}(1+e^{-i\varphi})\right|^2 = \cos^2\frac{\varphi}{2}$ and in port 5 as $I_5^I(\varphi) = \left|\frac{1}{2}(e^{-i\varphi}-1)\right|^2 = \sin^2\frac{\varphi}{2}$. The photocurrents generated by the correspondent PD's are directly proportional to these intensities: $j_4^I(\varphi) \sim I_4^I(\varphi)$ and $j_5^I(\varphi) \sim I_5^I(\varphi)$, so the PBD's output photocurrent is: $j_R^I(\varphi) = j_5^I(\varphi) - j_4^I(\varphi)$. In turn, the voltage oscillation that is observed on oscilloscope is directly proportional to the generated photocurrent. The actual proportionality constant is a function of: fraction of the beam area covering the PD's surface; PD's responsivity; LP's attenuation in transmission axis as well as impedance of an oscilloscope. Exactness of the constant estimation is, however, of no essence for this experiment as it scales equally for both polarization modes and we are only comparing the relative scaling factor for different configurations. Consequently, we represent the modeled voltage amplitude in arbitrary units in range $[-1,1]$ as a function of the photocurrent, i.e.: $V(j_R^I(\varphi)) = I_5^I(\varphi) - I_4^I(\varphi)$. Fig. 4 shows the plot of $V(j_R^I(\varphi))$ values out of full phase cycle $\varphi \in [0,2\pi]$:

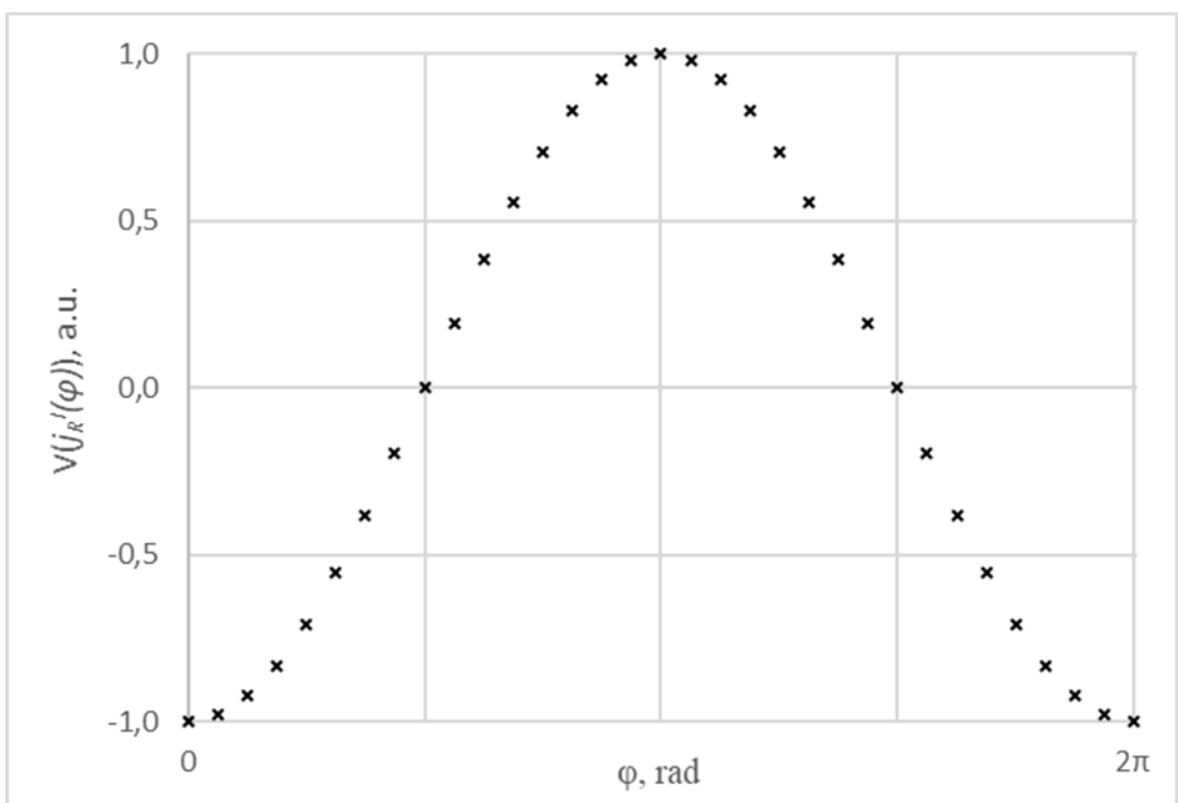

Fig. 4. Model of the voltage response in the test configuration 1.

Likewise, we model voltage of test configuration 2 the output state yields:

$$|\psi_{out}^{II}\rangle \rightarrow \frac{1}{2}\begin{pmatrix} 0.73(1-e^{-i\varphi})\widehat{D}_{4,v}(\alpha_{4,v}) \\ 0.73(1+e^{-i\varphi})\widehat{D}_{5,v}(\alpha_{5,v}) \\ 0 \\ 0 \end{pmatrix}|0\rangle \qquad (27)$$

The resulting voltage $V(j_R^{II}(\varphi))$ oscillation plot is shown on Fig. 5.

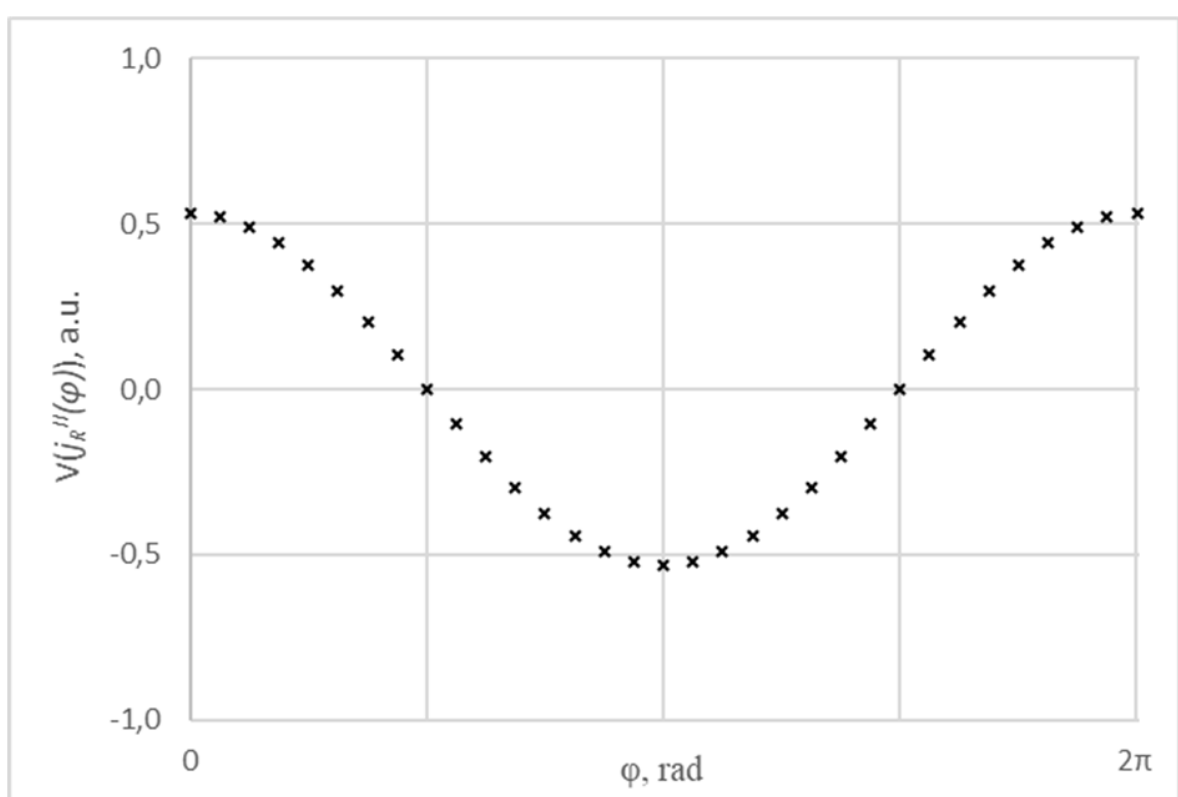

Fig. 5. Model of the voltage response in the test configuration 2.

In the third configuration we have:

$$|\psi_{out}^{III}\rangle \rightarrow \frac{1}{4}\begin{pmatrix} 0.73(1+e^{-i\varphi})\widehat{D}_{4,\mathrm{v}}(\alpha_{4,v}) \\ 0.73(1+e^{-i\varphi})\widehat{D}_{5,v}(\alpha_{5,v}) \\ (1+e^{-i\varphi})\widehat{D}_{4,h}(\alpha_{4,h}) \\ (1+e^{-i\varphi})\widehat{D}_{5,h}(\alpha_{5,h}) \end{pmatrix}|0\rangle \qquad (30)$$

where we can see the factor 0.73 cannot be omitted as it affects relative amplitudes. So, the oscillation plot for voltage $V(j_R^{III}(\varphi))$, Fig. 6. is:

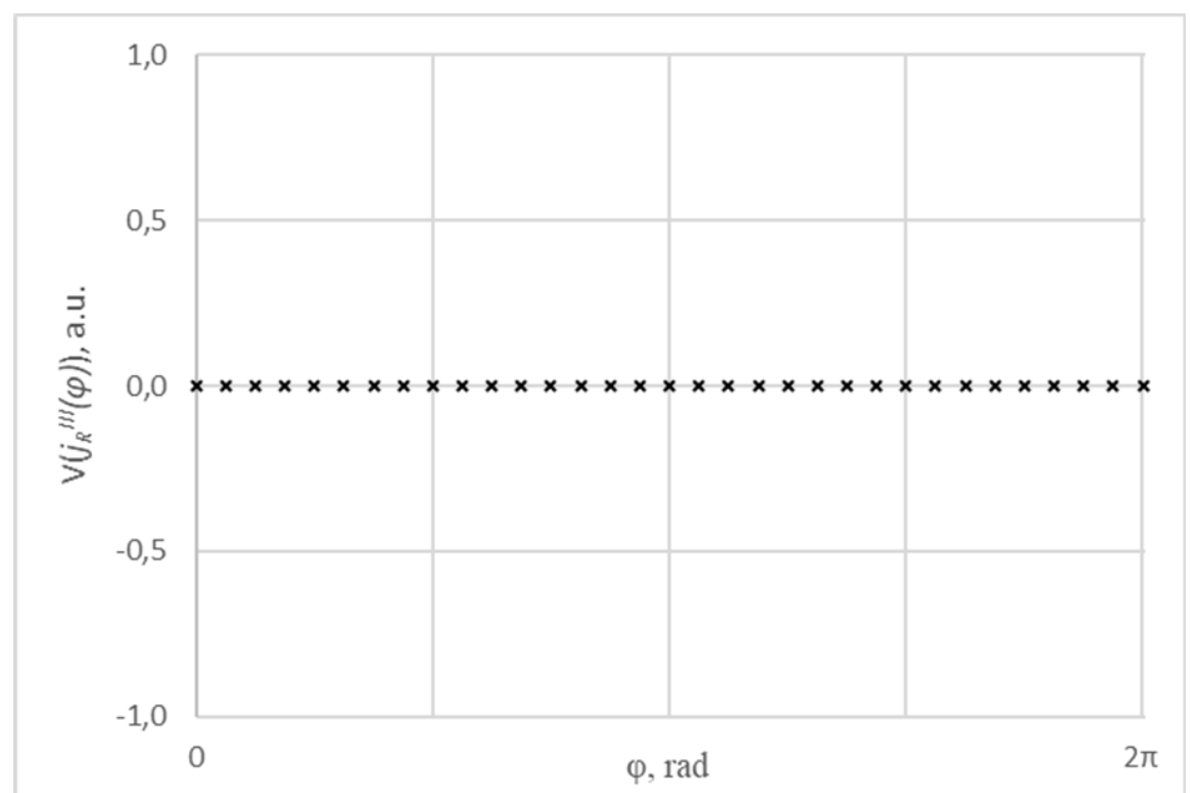

Fig. 6. Model of the voltage response in the test configuration 3.

In the same manner test configuration 4 output state becomes:

$$|\psi_{out}^{IIII}\rangle \rightarrow \frac{1}{4}\begin{pmatrix} 0.73(1+e^{-i\varphi})\widehat{D}_{4,\mathrm{v}}(\alpha_{4,v}) \\ 0.73(1-e^{-i\varphi})\widehat{D}_{5,v}(\alpha_{5,v}) \\ (1+e^{-i\varphi})\widehat{D}_{4,h}(\alpha_{4,h}) \\ (-1+e^{-i\varphi})\widehat{D}_{5,h}(\alpha_{5,h}) \end{pmatrix}|0\rangle \qquad (31)$$

The voltage of $V(j_R^{IIII}(\varphi))$ gives the following plot, Fig. 7.

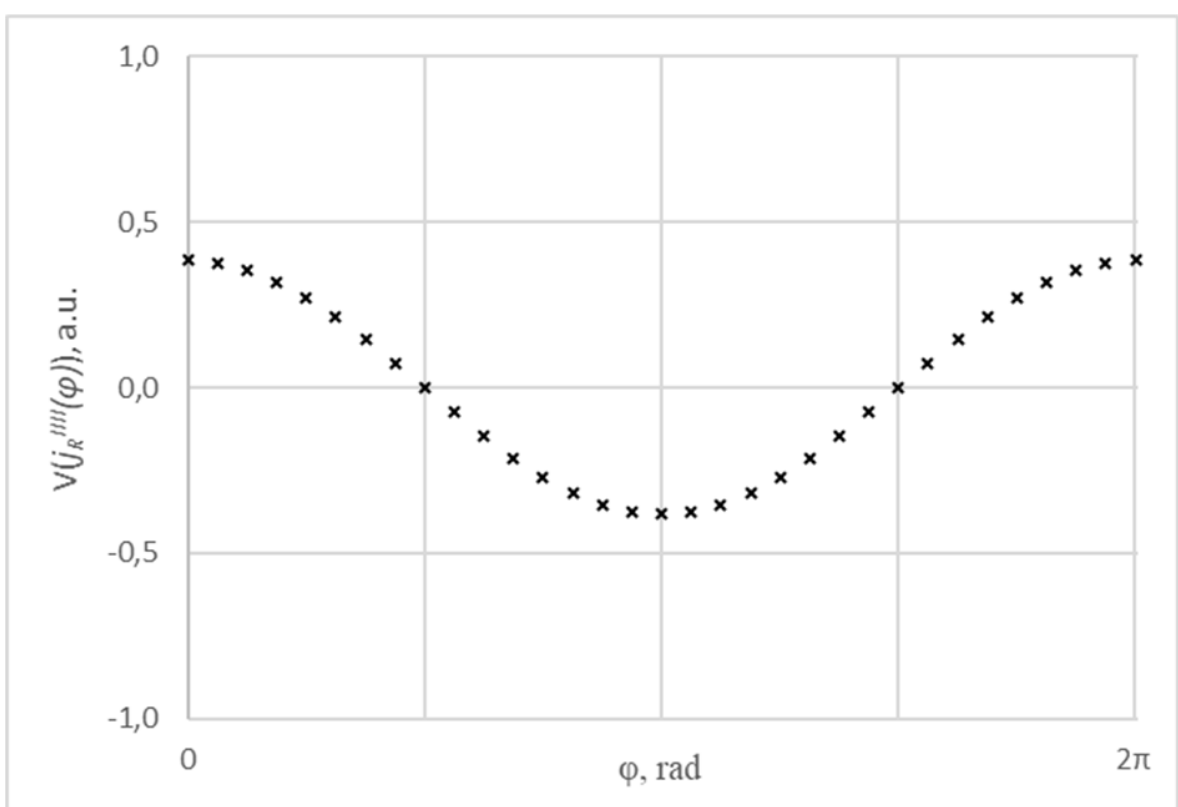

Fig. 7. Model of the voltage response in the test configuration 4.

The resulting output state for configuration 5 is:

$$|\psi_{out}^{V}\rangle \rightarrow \frac{1}{4}\begin{pmatrix} 0.73(-e^{-i\varphi}+1)\widehat{D}_{4,\mathrm{v}}(\alpha_{4,v}) \\ 0.73(e^{-i\varphi}-1)\widehat{D}_{5,v}(\alpha_{5,v}) \\ (-e^{-i\varphi}+1)\widehat{D}_{4,h}(\alpha_{4,h}) \\ (e^{-i\varphi}-1)\widehat{D}_{5,h}(\alpha_{5,h}) \end{pmatrix}|0\rangle \qquad (33)$$

Hence, the fifth configuration gives the following voltage response, Fig. 8.:

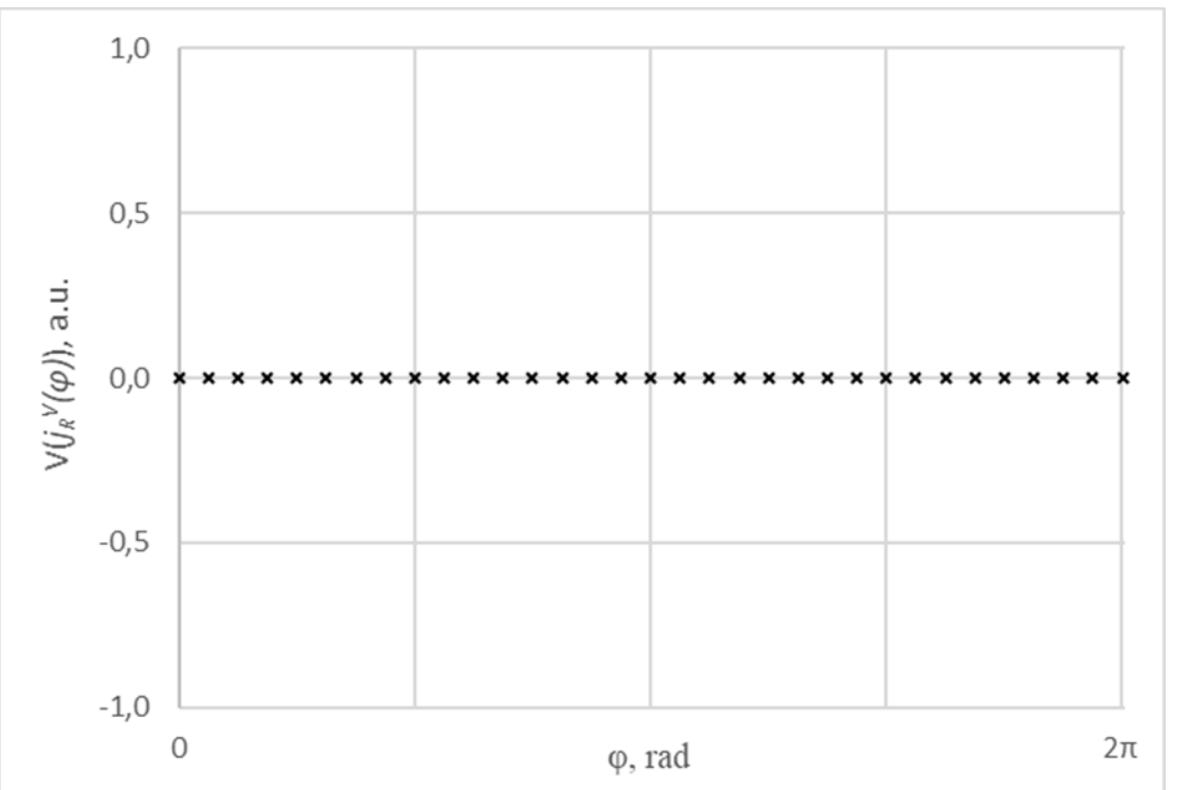

Fig. 8. Model of the voltage response in the test configuration 5.

Finally, in last configuration of the output state yields:

$$|\psi_{out}^{VI}\rangle \rightarrow \frac{1}{4}\begin{pmatrix} 0.73(-e^{-i\varphi}+1)\widehat{D}_{4,\mathrm{v}}(\alpha_{4,v}) \\ 0.73(e^{-i\varphi}+1)\widehat{D}_{5,v}(\alpha_{5,v}) \\ (-e^{-i\varphi}+1)\widehat{D}_{4,h}(\alpha_{4,h}) \\ (-e^{-i\varphi}-1)\widehat{D}_{5,h}(\alpha_{5,h}) \end{pmatrix}|0\rangle \qquad (34)$$

and the corresponding voltage model is shown on Fig. 9.:

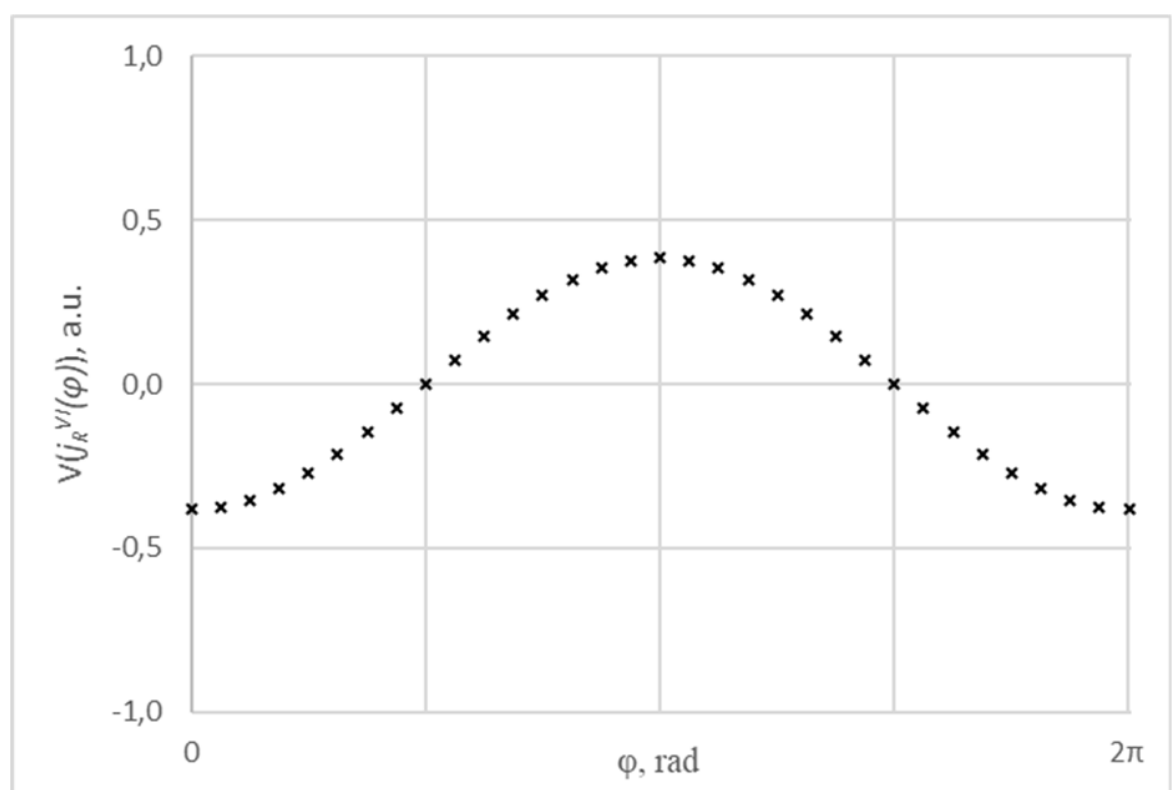

Fig. 9. Model of the voltage response in the test configuration 6.

We can now analyze an agreement of obtained models with respect to experimental data (Fig. 3.1-6.). This will require converting a.u. into volts. We begin with estimating the experimental error. The time series where no oscillation is modelled can be used to obtain an absolute value of error in measurement. Namely, we predicted 0 a.u. for any phase argument in test configurations 3 and 5, while experiment gives an oscillation depth of 56 mV and 28 mV in these tests respectively. Taking bigger value of 56 mV we see that signal deviates from 0 V at maximum by 36 mV. Thus, $\pm 36$ mV can be used as an absolute error in measurement. Coming back to conversion of a.u. into volts we see that experimentally the maximum amplitude of 1.52 V was observed with the test 1. This result corresponds to the maximum amplitude among our theoretical models in a.u. as well. Therefore, we take 1.52 V as a reference for a maximum $[-1,1]$ a.u. oscillation range and consequently the depth of $|-1|+|1|=2$ a.u. Hence, for the test configuration 2 we have the depth of oscillation $|I_5^{II}(0)-I_4^{II}(0)|+|I_5^{II}(\pi)-I_4^{II}(\pi)|=|0.5329|+|-0.5329|=1.0658$, whereas 1.0658 scales to 2 as 0.810 V (to 3 s.f.) scales to 1.52 V. We immediately see that predicted 0.810 V agrees with the experimentally measured value of 0.816 V $\pm$ 36 mV for test 2. In a similar manner, we obtain predicted values for all configurations and provide a summary for their agreement with experimental in the following Table II.

TABLE II.
VERIFICATION OF THEORETICAL AGREEMENT WITH THE EXPERIMENTAL DATA

| Test configuration | Theoretical prediction of the oscillation depth, a.u. (to 4 d.p.) | Theoretical prediction of the oscillation depth converted to volts with respect to 1.52 V, V (to 3 s.f.) | Experimental results of the oscillation depth with error in measurement, V (to 3 s.f.) |
|---|---|---|---|
| 1 | 2 | 1.520 | $1.520 \pm 0.036$ |
| 2 | 1.0658 | 0.810 | $0.816 \pm 0.036$ |
| 3 | 0 | 0 | $0.056 \pm 0.036$ |
| 4 | 0.7664 | 0.582 | $0.592 \pm 0.036$ |
| 5 | 0 | 0 | $0.028 \pm 0.036$ |
| 6 | 0.7664 | 0.582 | $0.572 \pm 0.036$ |

## V. CONCLUSIONS

We have experimentally verified the representation of 4-dimensional transformation matrix for a non-polarizing lossless beamsplitter. The experimental results agree with our theoretical predictions in each test apart from slight deviation in configuration 3 which amounts to about 3,5% relative error. This can be a result of a minor polarization selectivity of the test BS.

The results suggest that intensities in ports 4 and 5 oscillate either completely in phase or completely out of phase. They, therefore, feature a single phase permutation of $\pi$ rad. Provided there was an ideal transmission for all the modes we would not need to account for any additional scaling factors and obtain only three different levels of oscillations. Thus, we would have: the maximum level for tests 1 and 2; exactly half of the maximum level for tests 4 and 6 due to absorption of diagonal components in linear polarizer and no oscillation for tests 3 and 5. Taking into account just these three levels the results could have been interpreted qualitatively (in phase/out of phase). It is worth mentioning we also tested other types of beamsplitters such as plate (Thorlabs EBS1) and pellicle (Thorlabs BP150) one. They have different degree of polarization selectivity resulting only in different oscillation depth, yet the phase correlations were exactly the same as with the cube BS016. Therefore, the same phase distribution formalism holds for all these beamsplitters.

Obtaining the correct correlations even qualitatively involves accurate estimation of coefficients in the beamsplitter matrix. We showed that a common reciprocity approach for a lossless beamsplitter is insufficient for that. To give an actual example, the matrices:

$$\frac{1}{\sqrt{2}}\begin{pmatrix} 1 & i & 0 & 0 \\ i & 1 & 0 & 0 \\ 0 & 0 & 1 & i \\ 0 & 0 & i & 1 \end{pmatrix} \tag{35,a}$$

$$\frac{1}{\sqrt{2}}\begin{pmatrix} 1 & 1 & 0 & 0 \\ -1 & 1 & 0 & 0 \\ 0 & 0 & 1 & 1 \\ 0 & 0 & -1 & 1 \end{pmatrix} \tag{35,b}$$

satisfy the unitarity condition for a given 4x4 matrix but provide a wrong prediction for the correlations. Both (35,a) [5] and (35,b) predict an out of phase relation for test configurations 3 and 5, which is not supported by the experiment.

We proposed a full quantum optical BS matrix derivation by treating the input and output fields as single photon fields and imposing reciprocity and canonical commutation relations onto them. We showed that for a 4-dimensional BS correct matrix coefficients can be derived from commutator relation between the output ports. Moreover, this method is sufficient to immediately gain a full set of correct solutions and not just an additional restricting condition to the usual reciprocity relations. From the point of view of the rigorous formalism the technique is valid for coherent fields too. The displacement operators acting on the vacuum state can be interchanged with the Fock-state operators without any change to reflection and transmission amplitudes of the BS matrix.

At the same time we want to point out that it is a rather non-trivial phenomenon that exactly the same results can be obtained with absolutely different theoretical frameworks. On the one hand - Fresnel equations derived from the EM-wave boundary conditions at the macro-physical objects and on the other - commutation relations for the quantized fields where no wave interaction with a physical object is taking place.

In terms of a practical application the correct form of the 4x4 BS matrix can be used a 4-dimensional photonic Hadamard gate in the field of quantum computing. By implementing such Hadamard gates into linear quantum circuits, e.g. as discussed in [14] we expand total dimensions per component to four and hence practically obtain a 4-dimensional qubit. This actually quadruples processing power with respect to a 2-dimensional qubit.

## APPENDIX I

According to (9) we have:

$\hat{a}_s\hat{a}_{s'}^\dagger = 1 + \hat{a}_{s'}^\dagger\hat{a}_s$ if $s = s'$ and

$\hat{a}_s\hat{a}_{s'}^\dagger = \hat{a}_{s'}^\dagger\hat{a}_s$ if $s \neq s'$, hence:

$$\left[\hat{a}_c\hat{a}_v + \hat{a}_d\hat{a}_v, \hat{a}_c^\dagger\hat{a}_h^\dagger + \hat{a}_d^\dagger\hat{a}_h^\dagger\right] = 0 \Rightarrow$$

$$\left[\hat{a}_c\hat{a}_v, \hat{a}_c^\dagger\hat{a}_h^\dagger\right] + \left[\hat{a}_c\hat{a}_v, \hat{a}_d^\dagger\hat{a}_h^\dagger\right] + \left[\hat{a}_d\hat{a}_v, \hat{a}_c^\dagger\hat{a}_h^\dagger\right] + \left[\hat{a}_d\hat{a}_v, \hat{a}_d^\dagger\hat{a}_h^\dagger\right] = 0 \Rightarrow$$

solving the four commutators separately:

$$\left[\hat{a}_c\hat{a}_v, \hat{a}_c^\dagger\hat{a}_h^\dagger\right] = \left[t\hat{a}_a\hat{a}_v + r_v\hat{a}_b\hat{a}_v, t^*\hat{a}_a^\dagger\hat{a}_h^\dagger + r_h^*\hat{a}_b^\dagger\hat{a}_h^\dagger\right] = (t\hat{a}_a\hat{a}_v + r_v\hat{a}_b\hat{a}_v)\left(t^*\hat{a}_a^\dagger\hat{a}_h^\dagger + r_h^*\hat{a}_b^\dagger\hat{a}_h^\dagger\right) - \left(t^*\hat{a}_a^\dagger\hat{a}_h^\dagger + r_h^*\hat{a}_b^\dagger\hat{a}_h^\dagger\right)(t\hat{a}_a\hat{a}_v + r_v\hat{a}_b\hat{a}_v) = |t|^2\hat{a}_a\hat{a}_v\hat{a}_a^\dagger\hat{a}_h^\dagger + tr_h^*\hat{a}_a\hat{a}_v\hat{a}_b^\dagger\hat{a}_h^\dagger + r_vt^*\hat{a}_b\hat{a}_v\hat{a}_a^\dagger\hat{a}_h^\dagger + r_vr_h^*\hat{a}_b\hat{a}_v\hat{a}_b^\dagger\hat{a}_h^\dagger - |t|^2\hat{a}_a^\dagger\hat{a}_h^\dagger\hat{a}_a\hat{a}_v - t^*r_v\hat{a}_a^\dagger\hat{a}_h^\dagger\hat{a}_b\hat{a}_v - r_h^*t\hat{a}_b^\dagger\hat{a}_h^\dagger\hat{a}_a\hat{a}_v - r_h^*r_v\hat{a}_b^\dagger\hat{a}_h^\dagger\hat{a}_b\hat{a}_v = |t|^2\left(\hat{a}_a\hat{a}_a^\dagger\hat{a}_h^\dagger\hat{a}_v - \hat{a}_a^\dagger\hat{a}_a\hat{a}_h^\dagger\hat{a}_v\right) + tr_h^*\left(\hat{a}_b^\dagger\hat{a}_h^\dagger\hat{a}_a\hat{a}_v - \hat{a}_b^\dagger\hat{a}_h^\dagger\hat{a}_a\hat{a}_v\right) + r_vt^*\left(\hat{a}_a^\dagger\hat{a}_h^\dagger\hat{a}_b\hat{a}_v - \hat{a}_a^\dagger\hat{a}_h^\dagger\hat{a}_b\hat{a}_v\right) + r_vr_h^*\left(\hat{a}_b\hat{a}_b^\dagger\hat{a}_h^\dagger\hat{a}_v - \hat{a}_b^\dagger\hat{a}_b\hat{a}_h^\dagger\hat{a}_v\right) = |t|^2\left(\hat{a}_a\hat{a}_a^\dagger\hat{a}_h^\dagger\hat{a}_v - \hat{a}_a^\dagger\hat{a}_a\hat{a}_h^\dagger\hat{a}_v\right) + r_vr_h^*\left(\hat{a}_b\hat{a}_v\hat{a}_b^\dagger\hat{a}_h^\dagger - \hat{a}_b^\dagger\hat{a}_h^\dagger\hat{a}_b\hat{a}_v\right) = |t|^2\left(\left(1 + \hat{a}_a^\dagger\hat{a}_a\right)\hat{a}_h^\dagger\hat{a}_v - \hat{a}_a^\dagger\hat{a}_a\hat{a}_h^\dagger\hat{a}_v\right) + r_vr_h^*\left(\left(1 + \hat{a}_b^\dagger\hat{a}_b\right)\hat{a}_h^\dagger\hat{a}_v - \hat{a}_b^\dagger\hat{a}_b\hat{a}_h^\dagger\hat{a}_v\right) = (|t|^2 + r_vr_h^*)\hat{a}_h^\dagger\hat{a}_v;$$

$$\left[\hat{a}_c\hat{a}_v, \hat{a}_d^\dagger\hat{a}_h^\dagger\right] = \left[t\hat{a}_a\hat{a}_v + r_v\hat{a}_b\hat{a}_v, \tilde{r}_h^{\,*}\hat{a}_a^\dagger\hat{a}_h^\dagger + t^*\hat{a}_b^\dagger\hat{a}_h^\dagger\right] = (t\hat{a}_a\hat{a}_v + r_v\hat{a}_b\hat{a}_v)\left(\tilde{r}_h^{\,*}\hat{a}_a^\dagger\hat{a}_h^\dagger + t^*\hat{a}_b^\dagger\hat{a}_h^\dagger\right) - \left(\tilde{r}_h^{\,*}\hat{a}_a^\dagger\hat{a}_h^\dagger + t^*\hat{a}_b^\dagger\hat{a}_h^\dagger\right)(t\hat{a}_a\hat{a}_v + r_v\hat{a}_b\hat{a}_v) = t\tilde{r}_h^{\,*}\hat{a}_a\hat{a}_v\hat{a}_a^\dagger\hat{a}_h^\dagger + |t|^2\hat{a}_a\hat{a}_v\hat{a}_b^\dagger\hat{a}_h^\dagger + r_v\tilde{r}_h^{\,*}\hat{a}_b\hat{a}_v\hat{a}_a^\dagger\hat{a}_h^\dagger + r_vt^*\hat{a}_b\hat{a}_v\hat{a}_b^\dagger\hat{a}_h^\dagger - \tilde{r}_h^{\,*}t\hat{a}_a^\dagger\hat{a}_h^\dagger\hat{a}_a\hat{a}_v - \tilde{r}_h^{\,*}r_v\hat{a}_a^\dagger\hat{a}_h^\dagger\hat{a}_b\hat{a}_v - |t|^2\hat{a}_b^\dagger\hat{a}_h^\dagger\hat{a}_a\hat{a}_v - t^*r_v\hat{a}_b^\dagger\hat{a}_h^\dagger\hat{a}_b\hat{a}_v =$$

$$t\tilde{r}_h^{\,*}\left(\hat{a}_a\hat{a}_a^\dagger\hat{a}_h^\dagger\hat{a}_v - \hat{a}_a^\dagger\hat{a}_a\hat{a}_h^\dagger\hat{a}_v\right) + r_vt^*\left(\hat{a}_b\hat{a}_b^\dagger\hat{a}_h^\dagger\hat{a}_v - \hat{a}_b^\dagger\hat{a}_b\hat{a}_h^\dagger\hat{a}_v\right) = t\tilde{r}_h^{\,*}\left(\left(1 + \hat{a}_a^\dagger\hat{a}_a\right)\hat{a}_h^\dagger\hat{a}_v - \hat{a}_a^\dagger\hat{a}_a\hat{a}_h^\dagger\hat{a}_v\right) + r_vt^*\left(\left(1 + \hat{a}_b^\dagger\hat{a}_b\right)\hat{a}_h^\dagger\hat{a}_v - \hat{a}_b^\dagger\hat{a}_b\hat{a}_h^\dagger\hat{a}_v\right) = (t\tilde{r}_h^{\,*} + r_vt^*)\hat{a}_h^\dagger\hat{a}_v;$$

$$\left[\hat{a}_d\hat{a}_v, \hat{a}_c^\dagger\hat{a}_h^\dagger\right] = \left[\tilde{r}_v\hat{a}_a\hat{a}_v + t\hat{a}_b\hat{a}_v, t^*\hat{a}_a^\dagger\hat{a}_h^\dagger + r_h^*\hat{a}_b^\dagger\hat{a}_h^\dagger\right] = (\tilde{r}_v\hat{a}_a\hat{a}_v + t\hat{a}_b\hat{a}_v)\left(t^*\hat{a}_a^\dagger\hat{a}_h^\dagger + r_h^*\hat{a}_b^\dagger\hat{a}_h^\dagger\right) - \left(t^*\hat{a}_a^\dagger\hat{a}_h^\dagger + r_h^*\hat{a}_b^\dagger\hat{a}_h^\dagger\right)(\tilde{r}_v\hat{a}_a\hat{a}_v + t\hat{a}_b\hat{a}_v) = \tilde{r}_vt^*\hat{a}_a\hat{a}_v\hat{a}_a^\dagger\hat{a}_h^\dagger + \tilde{r}_vr_h^*\hat{a}_a\hat{a}_v\hat{a}_b^\dagger\hat{a}_h^\dagger + |t|^2\hat{a}_b\hat{a}_v\hat{a}_a^\dagger\hat{a}_h^\dagger + tr_h^*\hat{a}_b\hat{a}_v\hat{a}_b^\dagger\hat{a}_h^\dagger - t^*\tilde{r}_v\hat{a}_a^\dagger\hat{a}_h^\dagger\hat{a}_a\hat{a}_v - |t|^2\hat{a}_a^\dagger\hat{a}_h^\dagger\hat{a}_b\hat{a}_v - r_h^*\tilde{r}_v\hat{a}_b^\dagger\hat{a}_h^\dagger\hat{a}_a\hat{a}_v - r_h^*t\hat{a}_b^\dagger\hat{a}_h^\dagger\hat{a}_b\hat{a}_v =$$

$$\tilde{r}_vt^*\left(\hat{a}_a\hat{a}_a^\dagger\hat{a}_h^\dagger\hat{a}_v - \hat{a}_a^\dagger\hat{a}_a\hat{a}_h^\dagger\hat{a}_v\right) + tr_h^*\left(\hat{a}_b\hat{a}_b^\dagger\hat{a}_h^\dagger\hat{a}_v - \hat{a}_b^\dagger\hat{a}_h^\dagger\hat{a}_b\hat{a}_v\right) = \tilde{r}_vt^*\left(\left(1 + \hat{a}_a^\dagger\hat{a}_a\right)\hat{a}_h^\dagger\hat{a}_v - \hat{a}_a^\dagger\hat{a}_a\hat{a}_h^\dagger\hat{a}_v\right) + tr_h^*\left(\left(1 + \hat{a}_b^\dagger\hat{a}_h^\dagger\right)\hat{a}_h^\dagger\hat{a}_v - \hat{a}_b^\dagger\hat{a}_h^\dagger\hat{a}_b\hat{a}_v\right) = (\tilde{r}_vt^* + tr_h^*)\hat{a}_h^\dagger\hat{a}_v;$$

$$\left[\hat{a}_d\hat{a}_v, \hat{a}_d^\dagger\hat{a}_h^\dagger\right] = \left[\tilde{r}_v\hat{a}_a\hat{a}_v + t\hat{a}_b\hat{a}_v, \tilde{r}_h^{\,*}\hat{a}_a^\dagger\hat{a}_h^\dagger + t^*\hat{a}_b^\dagger\hat{a}_h^\dagger\right] = (\tilde{r}_v\hat{a}_a\hat{a}_v + t\hat{a}_b\hat{a}_v)\left(\tilde{r}_h^{\,*}\hat{a}_a^\dagger\hat{a}_h^\dagger + t^*\hat{a}_b^\dagger\hat{a}_h^\dagger\right) - \left(\tilde{r}_h^{\,*}\hat{a}_a^\dagger\hat{a}_h^\dagger + t^*\hat{a}_b^\dagger\hat{a}_h^\dagger\right)(\tilde{r}_v\hat{a}_a\hat{a}_v + t\hat{a}_b\hat{a}_v) = \tilde{r}_v\tilde{r}_h^{\,*}\hat{a}_a\hat{a}_v\hat{a}_a^\dagger\hat{a}_h^\dagger + \tilde{r}_vt^*\hat{a}_a\hat{a}_v\hat{a}_b^\dagger\hat{a}_h^\dagger + t\tilde{r}_h^{\,*}\hat{a}_b\hat{a}_v\hat{a}_a^\dagger\hat{a}_h^\dagger + |t|^2\hat{a}_b\hat{a}_v\hat{a}_b^\dagger\hat{a}_h^\dagger - \tilde{r}_h^{\,*}\tilde{r}_v\hat{a}_a^\dagger\hat{a}_h^\dagger\hat{a}_a\hat{a}_v - \tilde{r}_h^{\,*}t\hat{a}_a^\dagger\hat{a}_h^\dagger\hat{a}_b\hat{a}_v - t^*\tilde{r}_v\hat{a}_b^\dagger\hat{a}_h^\dagger\hat{a}_a\hat{a}_v - |t|^2\hat{a}_b^\dagger\hat{a}_h^\dagger\hat{a}_b\hat{a}_v =$$

$$\tilde{r}_v\tilde{r}_h^{\,*}\left(\hat{a}_a\hat{a}_a^\dagger\hat{a}_h^\dagger\hat{a}_v - \hat{a}_a^\dagger\hat{a}_a\hat{a}_h^\dagger\hat{a}_v\right) + |t|^2\left(\hat{a}_b^\dagger\hat{a}_b\hat{a}_h^\dagger\hat{a}_v - \hat{a}_b^\dagger\hat{a}_b\hat{a}_h^\dagger\hat{a}_v\right) = \tilde{r}_v\tilde{r}_h^{\,*}\left(\left(1 + \hat{a}_a^\dagger\hat{a}_a\right)\hat{a}_h^\dagger\hat{a}_v - \hat{a}_a^\dagger\hat{a}_a\hat{a}_h^\dagger\hat{a}_v\right) + |t|^2\left(\left(1 + \hat{a}_b^\dagger\hat{a}_b\right)\hat{a}_h^\dagger\hat{a}_v - \hat{a}_b^\dagger\hat{a}_b\hat{a}_h^\dagger\hat{a}_v\right) = (\tilde{r}_v\tilde{r}_h^{\,*} + |t|^2)\hat{a}_h^\dagger\hat{a}_v;$$

now summing over the commutators:

$$\left[\hat{a}_c\hat{a}_v, \hat{a}_c^\dagger\hat{a}_h^\dagger\right] + \left[\hat{a}_c\hat{a}_v, \hat{a}_d^\dagger\hat{a}_h^\dagger\right] + \left[\hat{a}_d\hat{a}_v, \hat{a}_c^\dagger\hat{a}_h^\dagger\right] + \left[\hat{a}_d\hat{a}_v, \hat{a}_d^\dagger\hat{a}_h^\dagger\right] = (|t|^2 + r_vr_h^*)\hat{a}_h^\dagger\hat{a}_v + (t\tilde{r}_h^{\,*} + r_vt^*)\hat{a}_h^\dagger\hat{a}_v + (\tilde{r}_vt^* + tr_h^*)\hat{a}_h^\dagger\hat{a}_v + (\tilde{r}_v\tilde{r}_h^{\,*} + |t|^2)\hat{a}_h^\dagger\hat{a}_v = (|t|^2 + r_vr_h^*)\hat{a}_h^\dagger\hat{a}_v + (t\tilde{r}_h^{\,*} + r_vt^* + \tilde{r}_vt^* + tr_h^*)\hat{a}_h^\dagger\hat{a}_v + (\tilde{r}_v\tilde{r}_h^{\,*} + |t|^2)\hat{a}_h^\dagger\hat{a}_v;$$

Thus:

$$\left[\hat{a}_c\hat{a}_v + \hat{a}_d\hat{a}_v, \hat{a}_c^\dagger\hat{a}_h^\dagger + \hat{a}_d^\dagger\hat{a}_h^\dagger\right] = 0 \Rightarrow$$

$$\left[\hat{a}_c\hat{a}_v, \hat{a}_c^\dagger\hat{a}_h^\dagger\right] + \left[\hat{a}_c\hat{a}_v, \hat{a}_d^\dagger\hat{a}_h^\dagger\right] + \left[\hat{a}_d\hat{a}_v, \hat{a}_c^\dagger\hat{a}_h^\dagger\right] + \left[\hat{a}_d\hat{a}_v, \hat{a}_d^\dagger\hat{a}_h^\dagger\right] = 0 \Rightarrow$$

$$|t|^2 + r_vr_h^* = 0;$$

$$t\tilde{r}_h^{\,*} + r_vt^* + \tilde{r}_vt^* + tr_h^* = 0;$$

$$\tilde{r}_v\tilde{r}_h^{\,*} + |t|^2 = 0.$$

## APPENDIX II

The detailed step by step derivation of an input state evolution according to the ports notations shown on Fig.2.:

The source is a single coherent input in port 1 representing a normalized vector state:

$$|\psi_{source}\rangle = \frac{1}{\sqrt{2}}\begin{pmatrix}\hat{D}_{1,v}(\alpha_{1,v})\\ 0\\ \hat{D}_{1,h}(\alpha_{1,h})\\ 0\end{pmatrix}|0\rangle \tag{13}$$

with $|0\rangle$ being a vacuum state, $\hat{D}_{1,v}(\alpha_{1,v})$ and $\hat{D}_{1,h}(\alpha_{1,h})$ being displacement operators [10] in spatial mode 1 in vertical and horizontal polarization modes accordingly. The action of PBS is represented by the matrix (14) which is CNOT gate in quantum computing [5]:

$$B_p = \begin{pmatrix}1 & 0 & 0 & 0\\ 0 & 1 & 0 & 0\\ 0 & 0 & 0 & 1\\ 0 & 0 & 1 & 0\end{pmatrix} \tag{14}$$

hence, yielding:

$$|\psi_{PBS}\rangle \rightarrow B_p|\psi_{source}\rangle \rightarrow \frac{1}{\sqrt{2}}\begin{pmatrix}\hat{D}_{2,v}(\alpha_{2,v})\\ 0\\ 0\\ \hat{D}_{3,h}(\alpha_{3,h})\end{pmatrix}|0\rangle \tag{15}$$

where $\hat{D}_{2,v}(\alpha_{2,v})$ and $\hat{D}_{3,h}(\alpha_{3,h})$ displacement operators in vertical and horizontal polarization modes shifted to spatial modes 2 and 3 accordingly. The phase modulation induced by the AOM is occurring in mode 3 and it is convenient to continue treating the diffracted beam which we selected for the recombination as spatial mode 3 as well. We derived the mode shifting for the diffracted beam using the function of phase modulation operator provided in [9], which adapted to our terms takes the following form:

$$\hat{S}_{PM} = \exp\left(\sum_1^\infty \hat{a}^\dagger_{\omega+\Omega} J_1(\omega)\hat{a}_\omega\right) \tag{16}$$

where: $\hat{a}_\omega$ is annihilation operator in the initial frequency mode of the source, $\hat{a}^\dagger_{\omega+\Omega}$ is creation operator in the AOM upshifted frequency mode and $J_1(\omega)$ is the first order Bessel function coefficient. It's worth adding a remark that expression of $\hat{S}_{PM}$ in the original paper contained a single-photon scattering coefficient instead of $J_1(\omega)$, however for a coherent state the expectation value of scattering coefficients summed over the infinity would plausibly result in $J_1(\omega)$. To show an action of the operator $\hat{S}_{PM}$ we represent the displacement operator $\hat{D}_{3,h}(\alpha_{3,h})$ explicitly in a single frequency mode $\omega$: $\hat{D}_{3,h,\omega}(\alpha_{3,h,\omega})$. Now we apply $\hat{S}_{PM}$ transformation suggested in [9] upon the displacement operator which shifts it to frequency mode $\omega + \Omega$:

$$\hat{S}_{PM}\hat{D}_{3,h,\omega}(\alpha_{3,h,\omega})\hat{S}^\dagger_{PM} = J_1(\omega)\hat{D}_{3,h,\omega+\Omega}(\alpha_{3,h,\omega+\Omega}) \tag{17}$$

Using the equation (17) we can represent the state evolution as follows:

$$|\psi_{AOM}\rangle \rightarrow \frac{1}{\sqrt{2}}\begin{pmatrix}\hat{D}_{2,v,\omega}(\alpha_{2,v,\omega})\\ \hat{S}_{PM}\hat{a}_{3,v,\omega}\hat{S}^\dagger_{PM}\\ \hat{a}_{2,h,\omega}\\ \hat{S}_{PM}\hat{D}_{3,h,\omega}(\alpha_{3,h,\omega})\hat{S}^\dagger_{PM}\end{pmatrix}|0\rangle \rightarrow$$
$$\frac{1}{\sqrt{2}}\begin{pmatrix}\hat{D}_{2,v,\omega}(\alpha_{2,v,\omega})\\ 0\\ 0\\ J_1(\omega)\hat{D}_{3,h,\omega+\Omega}(\alpha_{3,h,\omega+\Omega})\end{pmatrix}|0\rangle \tag{18}$$

where: $\hat{D}_{2,v,\omega}(\alpha_{2,v,\omega})$ is the displacement operator in vertical mode 2 and frequency $\omega$; $\hat{a}_{3,v,\omega}$ and $\hat{a}_{2,h,\omega}$ are annihilation operators in vertical and horizontal modes 3 and 2 and frequency $\omega$ accordingly. Adhering to formalities we have indicated the action of $\hat{S}_{PM}$ on the operator $\hat{a}_{3,v,\omega}$ as phase modulation applies to all operators in spatial mode 3. The resulting state $|\psi_{AOM}\rangle$ acquires different frequency modes and it is not normalized any longer due to $J_1(\omega)$ coefficient. Now, as discussed in previous chapter in the pre-measurement sequence step 2 by adjusting the HWP1 we equalize amplitudes for modes 2 and 3 and normalize the state accordingly obtaining:

$$|\psi'_{AOM}\rangle \rightarrow \frac{1}{\sqrt{2}}\begin{pmatrix}\hat{D}_{2,v,\omega}(\alpha_{2,v,\omega})\\ 0\\ 0\\ \hat{D}_{3,h,\omega+\Omega}(\alpha_{3,h,\omega+\Omega})\end{pmatrix}|0\rangle \tag{19}$$

The normalization is hence accounting for keeping the Bessel function coefficient $J_1(\omega)$ implicit throughout the derivation. It will also become apparent now why formalities with mode shifting were given in details when we consider the state in Schrödinger's picture, i.e. state $|\psi'_{AOM}\rangle$ evolving in time $\tau$ [11]:

$$|\psi'_{AOM}(\tau)\rangle \rightarrow \frac{1}{\sqrt{2}}\begin{pmatrix}\hat{D}_{2,v,\omega}\left(\alpha_{2,v,\omega}(\tau)\right)\\ 0\\ 0\\ \hat{D}_{3,h,\omega+\Omega}\left(\alpha_{3,h,\omega+\Omega}(\tau)\right)\end{pmatrix}|0\rangle \rightarrow$$
$$\frac{1}{\sqrt{2}}\begin{pmatrix}e^{-i\omega\tau}\hat{D}_{2,v,\omega}\left(\alpha_{2,v,\omega}(0)\right)\\ 0\\ 0\\ e^{-i(\omega+\Omega)\tau}\hat{D}_{3,h,\omega+\Omega}\left(\alpha_{3,h,\omega+\Omega}(0)\right)\end{pmatrix}|0\rangle \rightarrow$$
$$\frac{e^{-i\omega\tau}}{\sqrt{2}}\begin{pmatrix}\hat{D}_{2,v,\omega}\left(\alpha_{2,v,\omega}(0)\right)\\ 0\\ 0\\ e^{-i\Omega\tau}\hat{D}_{3,h,\omega+\Omega}\left(\alpha_{3,h,\omega+\Omega}(0)\right)\end{pmatrix}|0\rangle \tag{20}$$

It is apparent from the expression (20) that we deduced the phase factor $e^{-i\Omega\tau}$, which will affect the oscillation dynamics of an observable – light intensity. The state (20) can be simplified by omitting common phase factor $e^{-i\omega\tau}$ in future calculations. Additionally, we can treat the coherent states in modes $\omega$ and $\omega + \Omega$ as fully tangential, i.e. $\langle\alpha_\omega|\alpha_{\omega+\Omega}\rangle = 1$, since physically our detection equipment is equally sensitive to these modes within the actual shift of $\approx$107 MHz for $\Omega$. So, in further notations we can as well omit explicit indication of the frequency modes. Thus, we obtain the following normalized state:

$$|\psi'_{AOM}\rangle \to \frac{1}{\sqrt{2}}\begin{pmatrix} \widehat{D}_{2,v}(\alpha_{2,v}) \\ 0 \\ 0 \\ e^{-i\varphi}\widehat{D}_{3,h}(\alpha_{3,h}) \end{pmatrix}|0\rangle \tag{21}$$

with $\varphi$ being modulation phase $\varphi = \Omega\tau$. Using expression (21) we can define input states for test configurations. The action of HWP2 and HWP3 with FA set at $\frac{\pi}{4}$ is expressed in a matrix form using adapted Jones calculus definition [12], namely:

$$WP_2(\beta) = \begin{pmatrix} \cos 2\beta & 0 & \sin 2\beta & 0 \\ 0 & 1 & 0 & 0 \\ \sin 2\beta & 0 & \cos 2\beta & 0 \\ 0 & 0 & 0 & 1 \end{pmatrix} \tag{22}$$

for the HWP2 with $\beta$ being an angle of the wave-plate's FA and:

$$WP_3(\gamma) = \begin{pmatrix} 1 & 0 & 0 & 0 \\ 0 & \cos 2\gamma & 0 & \sin 2\gamma \\ 0 & 0 & 1 & 0 \\ 0 & \sin 2\gamma & 0 & \cos 2\gamma \end{pmatrix} \tag{23}$$

for the HWP3 with $\gamma$ being an angle of the wave-plate's FA.

$$|\psi^{I}_{in}\rangle \to WP_2\left(\frac{\pi}{4}\right)|\psi'_{AOM}\rangle \to \frac{1}{\sqrt{2}}\begin{pmatrix} 0 \\ 0 \\ \widehat{D}_{4,h}(\alpha_{4,h}) \\ e^{-i\varphi}\widehat{D}_{5,h}(\alpha_{5,h}) \end{pmatrix}|0\rangle \tag{24}$$

where $\widehat{D}_{4,h}(\alpha_{4,h})$ and $\widehat{D}_{5,h}(\alpha_{5,h})$ are displacement operators in horizontal polarization mode shifted to spatial modes 4 and 5 accordingly. Hence, in the first test configuration we have the following output state:

$$|\psi^{I}_{out}\rangle \to \frac{1}{2}\begin{pmatrix} 0 \\ 0 \\ (1+e^{-i\varphi})\widehat{D}_{4,h}(\alpha_{4,h}) \\ (e^{-i\varphi}-1)\widehat{D}_{5,h}(\alpha_{5,h}) \end{pmatrix}|0\rangle \tag{25}$$

Likewise, the input state of the test configuration 2 yields:

$$|\psi^{II}_{in}\rangle \to WP_3\left(\frac{\pi}{4}\right)|\psi'_{AOM}\rangle \to \frac{1}{\sqrt{2}}\begin{pmatrix} \widehat{D}_{4,v}(\alpha_{4,v}) \\ e^{-i\varphi}\widehat{D}_{5,v}(\alpha_{5,v}) \\ 0 \\ 0 \end{pmatrix}|0\rangle \tag{26}$$

So, taking into account the polarization selectivity of dispersive prisms P1 and P2 we obtain:

$$|\psi^{II}_{out}\rangle \to B|\psi^{II}_{in}\rangle \to \frac{1}{2}\begin{pmatrix} 0.73(1-e^{-i\varphi})\widehat{D}_{4,v}(\alpha_{4,v}) \\ 0.73(1+e^{-i\varphi})\widehat{D}_{5,v}(\alpha_{5,v}) \\ 0 \\ 0 \end{pmatrix}|0\rangle \tag{27}$$

where the $|\psi^{II}_{out}\rangle$ output state of the test configuration 2 is correlated to factor 0.73 so we can work in the same normalization scale for voltage response as in previous case. While one may suggest omitting the common factor 0.73 and consider the model normalization independently it may, however, appear confusing given the actual oscilloscope response. Thus, to have more quantitative approach we want to contrast the scales of all predicted models under the same normalization amplitudes. The factor 0.73 comes from $\sqrt{0.73}$ transmission amplitude for TE (vertical) mode per surface of dispersive prisms as discussed in experimental setup chapter.

In the third and fourth test configurations we use no waveplate but LP's with different angle settings. The effect of LP1 and LP2 can be defined according to the Jones calculus [12] as the following matrix suited for our 4-dimensional state vector:

$$LP(\theta_1,\theta_2) = \begin{pmatrix} \cos^2\theta_1 & 0 & \cos\theta_1\sin\theta_1 & 0 \\ 0 & \cos^2\theta_2 & 0 & \cos\theta_2\sin\theta_2 \\ \cos\theta_1\sin\theta_1 & 0 & \sin^2\theta_1 & 0 \\ 0 & \cos\theta_2\sin\theta_2 & 0 & \sin^2\theta_2 \end{pmatrix} \tag{28}$$

with $\theta_1$ and $\theta_2$ angles of transmission axes for LP1 and LP2 accordingly. Hence, in test configuration 3 we obtain:

$$|\psi^{III}_{in}\rangle \to |\psi'_{AOM}\rangle \tag{29}$$

$$|\psi^{III}_{out}\rangle \to LP\left(\frac{\pi}{4},\frac{\pi}{4}\right)B|\psi^{III}_{in}\rangle \to \frac{1}{4}\begin{pmatrix} 0.73(1+e^{-i\varphi})\widehat{D}_{4,v}(\alpha_{4,v}) \\ 0.73(1+e^{-i\varphi})\widehat{D}_{5,v}(\alpha_{5,v}) \\ (1+e^{-i\varphi})\widehat{D}_{4,h}(\alpha_{4,h}) \\ (1+e^{-i\varphi})\widehat{D}_{5,h}(\alpha_{5,h}) \end{pmatrix}|0\rangle \tag{30}$$

where we can see the factor 0.73 cannot be omitted as it affects relative amplitudes. The expression (30) shows that all of four amplitudes are in phase, so by subtracting the intensities at port 4 and 5 which now have two components per port: $I_4^{III}(\varphi) = I_5^{III}(\varphi) = \left|\frac{0.73}{4}(1+e^{-i\varphi})\right|^2 + \left|\frac{1}{4}(1+e^{-i\varphi})\right|^2 = \frac{1+0.73^2}{4}\cos^2\frac{\varphi}{2}$ and as they are both in phase subtracting the signals provides no oscillation. In the same manner test configuration 4 output state becomes:

$$|\psi^{IIII}_{out}\rangle \to LP\left(\frac{\pi}{4},-\frac{\pi}{4}\right)B|\psi^{IIII}_{in}\rangle \to \frac{1}{4}\begin{pmatrix} 0.73(1+e^{-i\varphi})\widehat{D}_{4,v}(\alpha_{4,v}) \\ 0.73(1-e^{-i\varphi})\widehat{D}_{5,v}(\alpha_{5,v}) \\ (1+e^{-i\varphi})\widehat{D}_{4,h}(\alpha_{4,h}) \\ (-1+e^{-i\varphi})\widehat{D}_{5,h}(\alpha_{5,h}) \end{pmatrix}|0\rangle \tag{31}$$

For the input vectors $|\psi^{V}_{in}\rangle$ and $|\psi^{VI}_{in}\rangle$, i.e. of last two test configurations, we perform a polarization state swap by applying both waveplate matrices upon vector $|\psi'_{AOM}\rangle$ (for which the order does not matter as these matrices commute):

$$\left|\psi_{in}^{V}\right\rangle = \left|\psi_{in}^{VI}\right\rangle \rightarrow WP_3\left(\frac{\pi}{4}\right)WP_2\left(\frac{\pi}{4}\right)|\psi'_{AOM}\rangle \rightarrow$$

$$\frac{1}{\sqrt{2}}\begin{pmatrix} 0 \\ e^{-i\varphi}\widehat{D}_{5,v}(\alpha_{5,v}) \\ \widehat{D}_{4,h}(\alpha_{4,h}) \\ 0 \end{pmatrix}|0\rangle \qquad (32)$$

Thus, resulting output state for configuration 5 is:

$$|\psi_{out}^{V}\rangle \rightarrow LP\left(\frac{\pi}{4},\frac{\pi}{4}\right)B\left|\psi_{in}^{V}\right\rangle \rightarrow$$

$$\frac{1}{4}\begin{pmatrix} 0.73(-e^{-i\varphi}+1)\widehat{D}_{4,\mathrm{v}}(\alpha_{4,v}) \\ 0.73(e^{-i\varphi}-1)\widehat{D}_{5,v}(\alpha_{5,v}) \\ (-e^{-i\varphi}+1)\widehat{D}_{4,h}(\alpha_{4,h}) \\ (e^{-i\varphi}-1)\widehat{D}_{5,h}(\alpha_{5,h}) \end{pmatrix}|0\rangle \qquad (33)$$

Finally, in last configuration of the output state we apply the anti-symmetric position of LP's:

$$|\psi_{out}^{VI}\rangle \rightarrow LP\left(\frac{\pi}{4},-\frac{\pi}{4}\right)B\left|\psi_{in}^{VI}\right\rangle \rightarrow$$

$$\frac{1}{4}\begin{pmatrix} 0.73(-e^{-i\varphi}+1)\widehat{D}_{4,\mathrm{v}}(\alpha_{4,v}) \\ 0.73(e^{-i\varphi}+1)\widehat{D}_{5,v}(\alpha_{5,v}) \\ (-e^{-i\varphi}+1)\widehat{D}_{4,h}(\alpha_{4,h}) \\ (-e^{-i\varphi}-1)\widehat{D}_{5,h}(\alpha_{5,h}) \end{pmatrix}|0\rangle \qquad (34)$$